\documentclass[
reprint,
superscriptaddress,
amsmath,
amssymb,
aps,
pra
]{revtex4-1}

\usepackage[T1]{fontenc} 

\usepackage{graphicx}
\usepackage{dcolumn}
\usepackage{bm}
\usepackage{lipsum}
\usepackage{color}
\usepackage[normalem]{ulem}
\usepackage{simplewick}
\usepackage{qcircuit}
\usepackage{braket}
\usepackage{float}
\usepackage{multirow}
\usepackage{tikz}
\usepackage{xr-hyper}
\usepackage[colorlinks=true,urlcolor=blue,citecolor=blue,linkcolor=blue]{hyperref}

\newcommand{\bF}{\bm F}
\newcommand{\bR}{\bm R}
\newcommand{\bv}{\bm v}

\begin{document}

\title{Microcanonical and finite temperature \textit{ab initio} molecular dynamics simulations on quantum computers}

\author{Igor O. Sokolov}
\affiliation{IBM Quantum, IBM Research -- Zurich, Switzerland}
\affiliation{Department of Chemistry, University of Zurich, Winterthurerstrasse 190, 8057 Zurich, Switzerland}

\author{Panagiotis Kl. Barkoutsos} 
\affiliation{IBM Quantum, IBM Research -- Zurich, Switzerland}

\author{Lukas Moeller} 
\affiliation{IBM Quantum, IBM Research -- Zurich, Switzerland}

\author{Philippe Suchsland} 
\affiliation{IBM Quantum, IBM Research -- Zurich, Switzerland}
\affiliation{Department of Physics, University of Zurich, Winterthurerstrasse 190, 8057 Zurich, Switzerland}

\author{Guglielmo Mazzola} 
\affiliation{IBM Quantum, IBM Research -- Zurich, Switzerland}

\author{Ivano Tavernelli}
\email{ita@zurich.ibm.com}
\affiliation{IBM Quantum, IBM Research -- Zurich, Switzerland}

\date{\today}

\begin{abstract}
\textit{Ab initio} molecular dynamics (AIMD) 
is a powerful tool to predict properties of molecular and condensed matter systems.
The quality of this procedure is based on accurate electronic structure calculations.
The development of quantum processors has shown great potential for the efficient evaluation of accurate ground and excited state energies of molecular systems, opening up new avenues for molecular dynamics simulations. 
In this work we address the use of variational quantum algorithms for the calculation of accurate atomic forces to be used in AIMD. 
In particular, we provide solutions for the alleviation of the statistical noise associated to the measurements of the expectation values of energies and forces, as well as schemes for the mitigation of the hardware noise sources (in particular, gate infidelities, qubit decoherence and readout errors).
Despite the relative large error in the calculation of the potential energy, our results show that the proposed algorithms can provide reliable MD trajectories in the microcanonical (constant energy) ensemble. 
Further, exploiting the intrinsic noise arising from the quantum measurement process, 
we also propose a Langevin dynamics algorithm for the simulation of canonical, i.e., constant temperature, dynamics.
Both algorithms (microcanonical and canonical) are applied to the simulation of simple molecular systems such as $\rm{H_2}$ and $\rm{H_3^+}$.
Finally, we also provide results for the dynamics of $\rm{H_2}$ obtained with IBM quantum computer \textit{ibmq\_athens}.
\end{abstract}

\keywords{Molecular Dynamics, Langevin dynamics, Geometry Optimization, Vibrational Spectrum, Hardware Efficient Ansatz, Energy Derivatives, Forces, VQE, Quantum Computing}

\maketitle

\section{\label{sec:intro}Introduction}

Quantum computing is emerging as a new computational paradigm for the solution, among others, of quantum mechanical many-body problems. 
In particular, in the recent years we have witnessed a rapid development of quantum algorithms in electronic structure calculations of both ground and excited states properties~\cite{Moll2018, cao2019quantum, mcardle2020quantum, ollitrault2020hardware}.
The Variational Quantum Eigensolver (VQE) algorithm~\cite{peruzzo_variational_2014} allows for the efficient calculation of the electronic structure of simple molecules both in simulations~\cite{Barkoutsos2018, sokolov2020quantum}
as well as on hardware experiments~\cite{kandala2017hardware, Ganzhorn2018, Omalley2017, Hempel2018, Santagati2018, Colless2018, Kandala2019, Arute2020_HF}. 
Typically, in the VQE algorithm, the molecular wavefunction is encoded in the quantum register using a classically inspired wavefunction Ansatz, such as unitary coupled cluster~\cite{Omalley2017,shen2017quantum}, 
or by means of an `heuristic' expansion dictated by the available gates and connectivity in the hardware~\cite{Barkoutsos2018}.

In all cases it is possible to show that a polynomial number of variational parameters (qubit rotations) are sufficient to achieve accurate results within the so-called chemical accuracy (i.e., with an error less than $1$ kcal/mol or $1.6\cdot10^{-3}$ Hartree).
Of particular relevance is the theoretical scaling of these new algorithms, which can achieve an accurate solution of the many-electron Schr\"odinger equation for molecular and solid state systems with a ${O}(N_{b}^{4})$ scaling in the number of basis functions, $N_b$. 
It is worth mentioning, that despite the potential advantage of the quantum algorithms, the simulation of the quantum circuits is memory limited on classical computers due to the exponential scaling of the Hilbert space in the number of qubits, while the state-of-the-art quantum hardware is still too noisy for achieving chemical accuracy on shallow circuits. 

In addition to the calculation of energies, quantum algorithms can provide an efficient solution to the calculation of the \textit{ab-initio} forces on the classical, point-like, atomic particles. 
These are of particular importance for the calculation of optimized molecular structure (annealing) as well as to perform molecular dynamics (MD) in the different thermodynamic ensembles.
In fact, the quantum circuit optimized for the calculation of the ground state energy can also be used for the evaluation of the expectation values of the gradient of the Hamiltonian with respect to the nuclear coordinates, giving access to the nuclear forces.
Several approaches have been already described in the literature. 
The most direct approach relies on finite difference (FD) approximations~\cite{milne2000calculus}, which can lead to fairly accurate forces at the cost of $6N$ additional wavefunction optimizations with $N$ being the number of nuclei.
To reduce the overhead, one can also obtain accurate gradients with a modified FD form that only requires a single wavefunction evaluation as shown in Ref.~\cite{kassal2009quantum}; this will also be the method of choice in this work. 
Alternatively, one can also use analytic gradients in the framework of VQE~\cite{mitarai2020theory} or the Lagrangian formalism, which allows the determination of the response functions with respect to the nuclear displacement without the explicit calculation of the derivatives~\cite{parrish2019hybrid}. \\

In this work, we are investigating quantum algorithms for the calculation of atomic forces to perform molecular dynamics in the microcanonical NVE ensemble (i.e. constant number of particles, volume and energy) as well as in the constant temperature canonical NVT ensemble (i.e. constant number of particles, volume and temperature). 
In the first case, we use well-studied classical integration scheme of the Newton's equations of motion, namely the Verlet algorithm~\cite{allen_computer_1987} to compute constant energy trajectories using VQE energies and forces. 
In the second case, we make use of the intrinsic statistical noise in evaluating quantum observables in a quantum computer using a generalized Langevin dynamics~\cite{Attaccalite08,kuhne_efficient_2007} scheme at constant (non-zero) temperature.
Both approaches are applied to the simulation of the dynamics of simple molecular systems such as H$_2$ and H$_3^{+}$.

The dynamics is performed using both the matrix representation (MR) of the parametrized VQE circuit as well as the classical emulation of the VQE algorithm, as it would be executed on a quantum computer. 
In this case, a realistic representation of the hardware noise, including fidelity of the qubit operations, qubit decoherence and readout errors, is applied to mimic as close as possible the hardware conditions. 
Finally, for the H$_2$ molecule, we also perform a study of the dynamics in the NVE ensemble, to demonstrate the feasibility but also the current limitations of our approach on current noisy hardware.

The main goal of this work remains the study of quantum algorithms for the calculation of forces in the presence of realistic noise models. 
In particular, we focus on the requirements in terms of qubit fidelity, decoherence and measurement error rates necessary to obtain reliable trajectories for near-term quantum calculations. 
We investigate the impact of the noise on the determination of the system energy and on the direct evaluation of the forces, showing evidences for a stable MD scheme based on a modified FD approach that, while approximated, can lead to stable dynamics. 
Due to the dominant role of the noise, we do not explore in this work other more sophisticated approaches~\cite{temme2017error, Ying2017, Endo2017, dimitrescu2018, Kandala2019, suchsland2020},
which improve the accuracy of the formal derivation of the forces at the cost of increasing the computational requirements.
However, to mitigate the impact of the noise, we also use an error mitigation scheme based on the Lanczos algorithm~\cite{lanczos1950iteration} and power iteration method~\cite{mises1929praktische} that improves substantially the quality of the energies and forces without modifying the overall scaling of the MD algorithm.

This paper is organized as follows.
In Section~\ref{sec:theory}, we define the molecular Hamiltonian and discuss re-usability of measurement outcomes of energy expectation value for forces.
We define the forces computed with FD and the Hellmann-Feynman theorem followed by the types of noise that influence their expectation values.
To mitigate the effect of noise, Lanczos method is introduced.
In Section~\ref{sec:md_theory}, we discuss the MD shemes for microcanonical dynamics using Verlet algorithm and, for canonical ensembles, the generalized Langevin dynamics that exploits the statistical noise associated to the quantum measurements process.
The results are presented in Section~\ref{sec:results} where we first show the outcomes of the geometry optimization and then, the microcanonical MD simulations of H$_2$ and H$_{3}^{+}$ molecules.
Finally, the Langevin dynamics simulations demonstrate how the canonical distribution (Boltzmann) can also be achieved.
Conclusions are presented in Section~\ref{sec:conclusion}.

\section{\label{sec:theory} Calculation of forces}

\subsection{Electronic ground state calculation}
\label{ss:gs}

The basis of our approach is the Born-Oppenheimer approximation~\cite{born1927quantentheorie}, which allows us to separate the dynamics of the electronic and nuclear subsystems.
More specifically, we constrain the ionic dynamics to adiabatically follow the potential energy surface (PES) defined by solution of the instantaneous electronic ground state at fixed ionic positions.

As standard practice in state-of-the-art electronic structure calculations in quantum computers~\cite{Moll2018, cao2019quantum, mcardle2020quantum}, we adopt the second quantized approach to solve the molecular Hamiltonian
\begin{align}
\hat{H} (\bm{R}) &=\sum_{rs} h_{rs}(\bm{R})\, \hat{a}^{\dagger}_{r} \hat{a}_{s} \label{eq:H}  \\ &+
\frac{1}{2}\sum_{rstu} g_{rstu}(\bm{R}) \,   \hat{a}^{\dagger}_{r} \hat{a}^{\dagger}_{s} \hat{a}_{u} \hat{a}_{t} + E_{NN}(\bm{R}), \notag
\end{align}
with $h_{rs}(\bm{R})$ denoting the one-electron integrals and $g_{rstu}(\bm{R})$ the two-electron integrals in physics notation that are commonly obtained with a Hartree-Fock (HF) calculation.
Notice that the collective vector of nuclear coordinates $\bm{R}=$ ($\bm{R}_1$,$\bm{R}_2$,...,$\bm{R}_N$) of $N$ nuclei in $\mathbb{R}^{3N}$ simply parametrizes the electronic Hamiltonian. 
The position vector of a single nuclei $I\in\{1,...,N\}$ is denoted by $\bm{R}_I=({R}_{Ix}, {R}_{Iy}, {R}_{Iz})$.
The operators $\hat{a}_{r}^{\dagger}$~($\hat{a}_{r}$) represent the fermionic creation~(annihilation) operators for electrons in 
HF spin-orbitals (MOs).
The indices $r,s,t,u$ are used to label general (occupied or virtual) MOs.
The term $E_{NN}(\bm{R})$ describes the nuclear repulsion energy.
The full form of terms in Eq.~\eqref{eq:H} is given in Appendix~\ref{sec:appendixD_integrals}.

The enabling step of the approach is to find first the ground state of Eq.~\eqref{eq:H}. 
While in general this task is not achievable exactly, a close approximation of the ground state can be obtained variationally by using the VQE algorithm.
This method features quantum circuits with gates that are defined collectively by optimizable parameters $\bm{\theta}$.
This generates a parametrized quantum state $ | \Psi (\bm{\theta}) \rangle$, often called \emph{trial state}.
These parameters are optimized to minimize the energy $E(\bm{\theta}) = \langle \Psi (\bm{\theta})| \hat{H}(\bm{R})| \Psi (\bm{\theta})\rangle$ for a given Hamiltonian. 
This optimization is performed \emph{classically}.
Since this approach is a well-established framework, we direct the reader to Refs.~\cite{peruzzo_variational_2014, mcclean_theory_2016}.

One point that we want to stress here concerns the practical evaluation of the energy as well as the expectation value
\begin{equation}
\label{e:expect}
  \langle \hat{\mathcal{O}}  \rangle  =  \langle \Psi (\bm{\theta})| \hat{\mathcal{O}}| \Psi (\bm{\theta})\rangle
\end{equation}
of any other operator $\hat{\mathcal{O}}$.
Formally, any hermitian operator $\hat{\mathcal{O}}$ defined on an $n$-qubit Hilbert space can be decomposed as
\begin{equation}
    \label{eq:paulis_decomposition}
    \hat{\mathcal{O}} = \sum_{\lambda=1}^{\Lambda} c_\lambda ~\hat{P}_\lambda\, , \hspace{1cm} c_\lambda \in \mathbb{R} \, .
\end{equation}
Each of the $\Lambda$  $n$-qubit Pauli strings is an element of the set $\mathcal{P}_n = \lbrace \hat{p}_1 \otimes
\hat{p}_2 \otimes \dots \otimes \hat{p}_n~|~\hat{p}_i \in \lbrace \hat{I}, \hat{X}, \hat{Y}, \hat{Z} \rbrace \rbrace$, and is tensor products of $n$ single qubit Pauli operators (see Appendix~\ref{sec:appendixF_paulimat}). 

The expectation value in Eq.~\eqref{e:expect}
is calculated as the sum of the  expectation values $ \langle \hat{P}_\lambda \rangle$ of the single Pauli operator, multiplied by the respective scalar number $c_\lambda$.
Finally, the expectation value $ \langle \hat{P}_\lambda \rangle$ can  be obtained by sampling from the prepared state $ |\Psi \rangle$ using $\mathcal{N}$  measurements, hence $\mathcal{N}$ repetition of the same circuit (see~\cite{kandala2017hardware} for details).
The statistical error associated with the evaluation of $ \langle \hat{P}_\lambda \rangle$ decreases as $1 / \sqrt{\mathcal{N}}$.
The total number of measurements required to compute $ \langle \hat{\mathcal{O}}  \rangle$ is therefore $\Lambda \mathcal{N}$, assuming for simplicity to allocate the same number of resources for each Pauli string.
While extensive algorithmic efforts have been recently put forward to mitigate this issue \cite{kandala2017hardware,torlai2020precise,kubler2020adaptive,yen2020measuring,izmaylov2019unitary,crawford2019efficient}, the impact of the statistical noise in evaluating quantum observables remains a peculiar aspect of quantum computation.

In the case of the energy, the target qubit operator $\hat{\mathcal{O}}$ is the Hamiltonian of Eq.~\eqref{eq:H} followed by a Jordan-Wigner transformation~\cite{Jordan1928} (for instance) that maps the fermionic operators $\hat{a}_{r}^{\dagger}$ and $\hat{a}_{r}$ to qubit operators, and the size of the qubit register $n$, is defined by the maximum number of molecular orbitals considered in Eq.~\eqref{eq:H}.
 
Suppose also that we want to measure not just one but several, say $3N$, operators $\hat{\mathcal{O}}_i$ which share the same support in the set $\mathcal{P}_n$ and only differ in the values  $\lbrace c_\lambda^{(i)} \rbrace$.
These $3N$ operators can therefore  be computed from the same dataset of $\Lambda \mathcal{N}$ circuit repetitions.
The expectation values of these operators become correlated and a covariance matrix is defined as
\begin{equation}\label{eq:covmatgen}
\textrm{Cov}_{ij} =  \left< ~
    \left(  \hat{\mathcal{O}}_i - \left< \hat{\mathcal{O}}_i \right> \right)
    \left(  \hat{\mathcal{O}}_j - \left< \hat{\mathcal{O}}_j \right> \right)
  ~\right>.
\end{equation}
This concept will become useful in the following discussion (see Section~\ref{ssc:Langevin}) where the covariance matrix is built from the force operators (Eq.~\eqref{eq:covmat}) and used to drive the molecular dynamics in the canonical ensemble.

\subsection{Force estimator}

The forces on the nuclei $\bm{F}=$ ($\bm{F}_1$,$\bm{F}_2$,...,$\bm{F}_N$) with $\bm{F}_I = ({F}_{Ix},{F}_{Iy},{F}_{Iz})$ are the derivatives of the energy with respect to the nuclear coordinates, $F_{I\alpha}(\bm{R}) = \left.\frac{\mathrm{d} E}{\mathrm{d}  R_{\mathrm{I} \alpha}}\right|_{\bm{R}}$, where $\alpha \in \{x, y, z\}$. 
The total derivative is given explicitly by
\begin{align}
\label{eq:forces} 
F_{I\alpha}(\bm{R}) & = 
\langle \Psi (\bm{\theta})| \partial_{I\alpha} \hat{H}(\bm{R})| \Psi (\bm{\theta})\rangle   \notag \\
&+ \langle \partial_{I\alpha}\Psi (\bm{\theta})| \hat{H}(\bm{R})| \Psi (\bm{\theta})\rangle \\
&+ \langle \Psi (\bm{\theta})|  \hat{H}(\bm{R})| \partial_{I\alpha} \Psi (\bm{\theta})\rangle \, , \notag
\end{align}
with $\partial_{I\alpha}=\frac{\partial}{\partial R_{\mathrm{I} \alpha}}$, where the second and last terms denote the wavefunction contributions, the so-called Pulay forces~\cite{pulay1969ab} (see Ref.~\cite{mcardle2019variational} for their possible implementation in a quantum circuit).
The first term corresponds to the Hellmann-Feynman force~\cite{guttinger1932verhalten,pauli1933principles,hellman1937einfuhrung,feynman1939forces}.

For an approximation of this derivative, the centered FD method can be applied as
\begin{align}
F_{I\alpha}^{\text{FD}}(\bm{R}) & = \frac{\langle\Psi_{+}|\hat{H}_{+}| \Psi_{+}\rangle-\langle\Psi_{-}|\hat{H}_{-}| \Psi_{-}\rangle}{2 \Delta R}, 
\label{eq:forces_FD}
\end{align}
where $\hat{H}_{\pm} = \hat{H}(\bm{R}\pm\Delta R \bm{e}_{I\alpha})$ with corresponding ground states $|\Psi_{\pm} \rangle$ respectively at $\bm{R}\pm\Delta R \bm{e}_{I\alpha}$,
with the unit vector $\bm{e}_{I\alpha}$ and the step-size $\Delta R$. 
In order to compute the forces $\bm{F}$, one needs to perform $6N$ electronic structure calculations (e.g. using VQE) to obtain the states $|\Psi_{\pm} \rangle$, making this approach computationally costly.
In addition, centered FD method typically introduces the numerical errors, i.e. round-off and discretization errors, with the latter scaling with the step-size as ${O}(\Delta R^2)$.
In the noisy setting, this approach suffers from large errors due to the presence of independent statistical errors in the computations of the left and right expectation values in Eq.~\eqref{eq:forces_FD}. 
That is alone sufficient to prevent its practical implementation~\cite{mitarai2020theory} which, at variance with quantum Monte Carlo methods, cannot be mitigated by the use of correlated sampling techniques~\cite{filippi2000correlated}.

Therefore, we opt for the Hellmann-Feynman approach which consists in considering only the first term in Eq.~\eqref{eq:forces} and applying numerical differentiation (FD) on $\partial_{I\alpha} \hat{H}(\bm{R})$ term as 

\begin{align}
F_{I\alpha}^{\text{H-F}}(\bm{R}) & =  \left\langle \Psi_0\left| \frac{\hat{H}_{+} - \hat{H}_{-}}{2\Delta R}\right| \Psi_0 \right\rangle,
\label{eq:forces_HF}
\end{align}
where $|\Psi_{0} \rangle$ is the optimized ground state wavefunction at geometry $\bm{R}$. 
The error due to this approximation is given by the Pulay force $\Delta F_{corr, I\alpha} = \langle \partial_{I\alpha}\Psi (\bm{\theta})| \hat{H}(\bm{R})| \Psi (\bm{\theta})\rangle + \langle \Psi (\bm{\theta})|  \hat{H}(\bm{R})| \partial_{I\alpha} \Psi (\bm{\theta})\rangle$ assuming that $\Delta R$ is chosen sufficiently small to minimize the contribution of the FD discretization error. 
In contrast with the standard FD approach presented in Eq.~\eqref{eq:forces_FD}, only a single wavefunction $|\Psi_{0} \rangle$ has to be obtained for $\bm{F}^{\text{H-F}}$ instead of $6N$ for $\bm{F}^{\text{FD}}$(i.e. by means of the VQE algorithm).

\subsection{Statistical and hardware noise sources}

In this section, we briefly summarize the different error types occurring in the implementation of the force algorithm on a quantum computer.
The first error source, that we name \emph{systematic}, stems from the approximate solution of the ground state electronic wavefunction for the Hamiltonian in Eq.~\eqref{eq:H}.
This error can be reduced systematically by employing more accurate trial wavefunctions, which can better represent the true ground state at fixed $\bR$, as well as by improving the (classical) optimization algorithm in the VQE approach.

The second error type originates from the statistical evaluation of the force operators of Eq.~\eqref{eq:forces_FD} with a finite number of measurements and we name it the \textit{statistical noise}.
This issue has been already discussed in Section~\ref{ss:gs} and cannot be alleviated even when using the exact ground state wavefunction.
However, in Section~\ref{ssc:Langevin} we will show how we can exploit statistical noise to perform finite temperature MD simulations.

Finally, the last source of errors is due to \textit{hardware noise}.
This is modelled by the gate, readout and thermalization errors represented by Kraus operators that are applied to the density matrix of the state~\cite{Nielse_Chuang2011}.
Gate and thermalization errors are introduced as depolarization and general amplitude damping channels.  
The gate and thermalization errors can be loosely categorized as systematic as they affect the possibility to reach the ground state.
The readout noise instead is hard to model and cannot easily be absorbed in the statistical noise contribution.
For a more detailed account on the implementation of these noise types in Qiskit~\cite{qiskit2019} see Appendix~\ref{sec:appendixB_noisesources}.

\subsection{\label{sec:forces_methods} Lanczos noise mitigation scheme}

A method to mitigate the noise of near-term quantum computers consists in applying the Lanczos mitigation scheme for the evaluation of the expectation value of operators in the VQE algorithm. 
The evaluation of the expectation value of a given operator $\hat{\mathcal{O}}$ with the VQE wavefunction $| \Psi_0\rangle$ is affected by hardware noise and measurement errors. 
Strictly speaking, this implies that, under noisy conditions, also higher `exact' eigenstates of the Hamiltonian $\hat{H}(\bm R)$ will contribute to the evaluation of the expectation value $\langle \Psi_0 |\hat{\mathcal{O}}|\Psi_0\rangle$.  
In order to alleviate this effect, we propose to use the Lanczos approach~\cite{suchsland2020, seki2020quantum}, which has the effect of partially projecting out the contributions of the excited states from the measurement of the expectation values, reducing therefore the effect of noise. 
When referring to the Lanczos mitigation scheme in the VQE-L algorithm, we will imply the substitution
\begin{align}
\langle \Psi_0 |\hat{\mathcal{O}}|\Psi_0\rangle \rightarrow &
\frac{\langle \Psi_0 | (\hat{H}(\bm R)-d)\hat{\mathcal{O}}(\hat{H}(\bm R)-d)| \Psi_0\rangle}{\langle \Psi_0 |(\hat{H}(\bm R)-d)^2 |\Psi_0\rangle} \nonumber \\
&=L_{d,\bm R}(\hat{\mathcal{O}}), \label{eq:lanczos1}
\end{align}
where $d \in \mathbb{R}$ is a tunable parameter that need to be optimized a priori. Note that the increase in accuracy is obtained at the cost of additional measurements of the terms in Eq.~\eqref{eq:lanczos1}, which include expectation values of $\hat{H}(\bm R) \hat{\mathcal{O}} \hat{H}(\bm R)$.

The method can be rationalized by interpreting $L_{d,\bm R}(\hat{\mathcal{O}})$ as the measurement of $\hat{\mathcal{O}}$ with respect to the modified state 
\begin{align}
   |\Psi_0\rangle \rightarrow \frac{(\hat{H}(\bm R) -d)|\Psi_0\rangle}{N_{\Psi_0}}=\sum_i \frac{(E_i(\bm R)-d) |E_i(\bm R)\rangle \alpha_i}{N_{\Psi_0}},
\end{align}
where we used the spectral representation of $\hat{H}(\bm R)$ in its eigenstates $\{|E_i(\bm R)\rangle\}$ with $|E_0(\bm R)\rangle$ the true ground state approximated by $| \Psi_0\rangle$, $N_{\Psi_0}^2=\langle \Psi_0 |(\hat{H}(\bm R)-d)^2 |\Psi_0\rangle$ and projection $\alpha_i=\langle E_i(\bm R)|\Psi_0\rangle$.
The coefficient associated to $|E_i(\bm R)\rangle$ is selected to fulfill the condition 
\begin{equation} \label{eq:lanczoe_condition}
\frac{|E_0(\bm R)-d|}{N_{\Psi_0}}>1,
\end{equation}
so that to increase the contribution of the ground state. 
In fact, by enforcing $|L_{d,\bm R}(\hat{H}(\bm R))-d|/N_{\Psi_0}>1$ and $d>L_{d,\bm R}(\hat{H}(\bm R))$ one gets that $|E_0(\bm{R})-d|/N_{\Psi_0}>1$~\cite{suchsland2020}.

The performance of Lanczos method depends therefore on the parameter $d$. In fact, while with the increase of the value of $d$ (within the limit defined in Eq.~\eqref{eq:lanczoe_condition}) the measurement confidence increases, 
the quality of the expectation values decreases. 
There is therefore a trade-off in the selection of $d$, which needs to be assessed independently for all observables of interest (see Fig.~\ref{H2_lanczos_d} in Appendix~\ref{sec:appendixE_lancz}).

\section{Molecular dynamics schemes}\label{sec:md_theory}

\begin{figure*}[ht]
\centering
\includegraphics[width=1.0\linewidth]{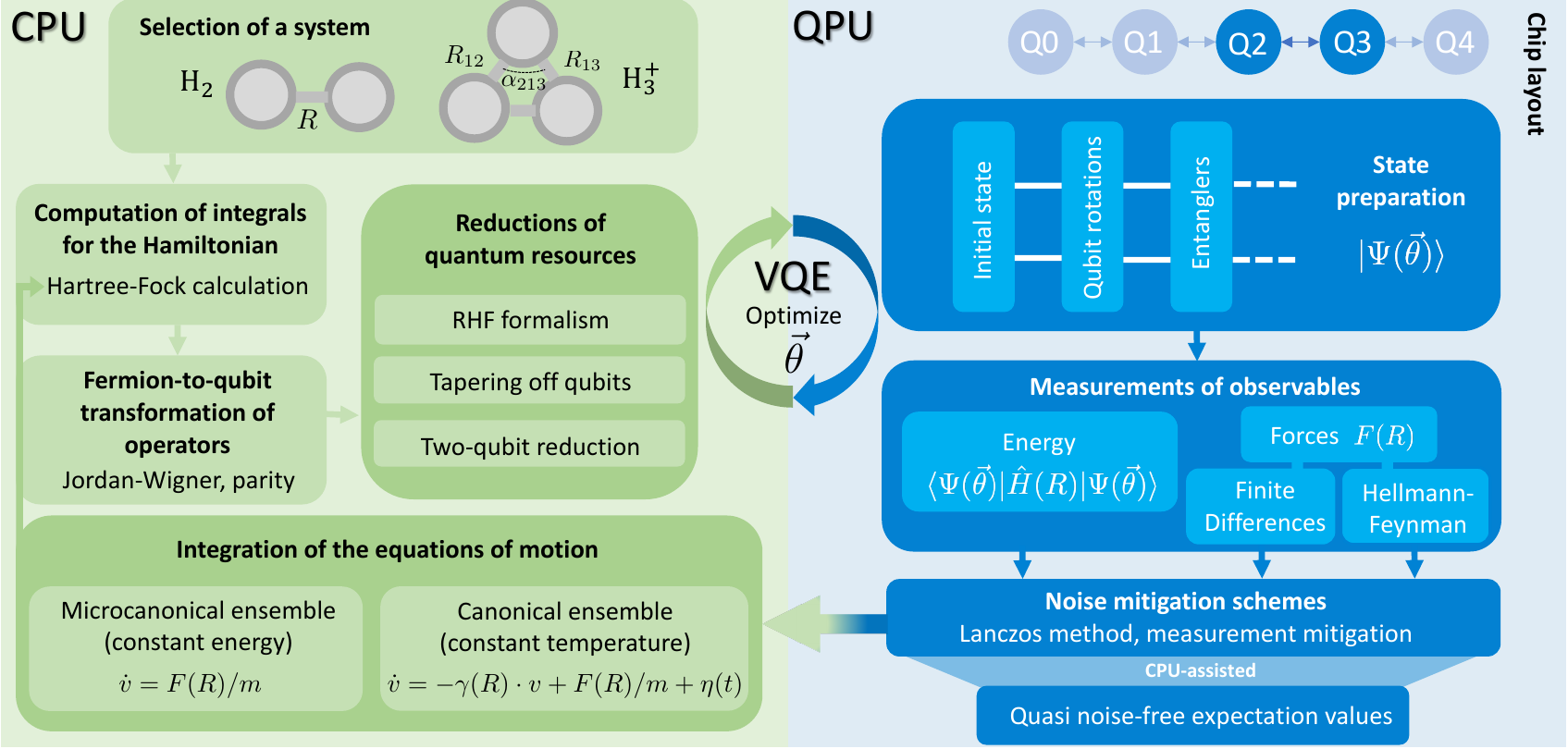}
\caption{{Hybrid quantum-classical approach for molecular dynamics simulations in second quantization. 
At each time step, the classical processing unit (CPU) performs a Hartree-Fock calculation for the evaluation of the molecular Hamiltonian and of the forces operators at the given molecular geometry.
The quantum processors (QPU) or a classical simulator of the QPU is then performing the optimization of the corresponding wavefunction (parametrized in the variables $\vec\theta$) and performs the evaluation of the required expectation values for the calculation of the energies and forces.
These values are then passed back to CPU, which performs the integration of the equations of motion in the chosen ensemble (microcanonical or canonical ensembles).
On the upper right corner, we show a sketch of the IBM Quantum device \textit{ibmq\_athens} used for the hardware calculations of the H$_2$ molecule (highlighted are the qubits used in the experiment and noisy simulations).
}}
	\label{fig:schema_md}
\end{figure*}

In this section, we present two molecular dynamics schemes for simulations in the microcanonical (isolated, NVE) and the canonical (thermalized, NVT or NPT) ensembles. 
In the first case, MD is performed at constant energy, and therefore a lot of care is necessary to reduce as much as possible the impact of the noise on the forces calculation.

In the second case, we will instead make use of the intrinsic noise of the quantum device to perform canonical MD simulations at constant temperature using Langevin dynamics. 
The tuning of the friction coefficient for the nuclear velocities will allow for the setting of the desired ensemble temperature.

\subsection{\label{sec:Verlet} Microcanonical dynamics using Verlet integrator}

Constant energy simulations can be straightforwardly achieved
by integrating Newton's equations of motion
\begin{eqnarray} 
\label{e:newton}
 \dot {\bv }    &=&       \bF(\bR) / {\bm m},   \\
 \dot{  \bR }   &=&  \bv,  \label{newton} 
\end{eqnarray}
where $\bv$ is the $3N$-dimensional vector made by the velocities of the $N$ nuclei, and ${\bm m}$ is a vector containing the masses of the $N$ particles~\cite{notemass}.
We employ the velocity-free Verlet integrator scheme~\cite{verlet1967computer},
\begin{equation}
    \bm{R}(t+\Delta t)=2\bm{R}(t)-\bm{R}(t-\Delta t)+\frac{1}{2}{\bm{F}}(t)/\bm{m}~\Delta t^{2}, \label{eq:verlet}
\end{equation}
that updates the positions of nuclei at the next time-step $t+\Delta t$, with an error ${O}(\Delta t^4)$.
The main issue with integrating the Newtonian equations of motion stems from the fact that
(\emph{i}) the \emph{systematic} error prevents the nuclei to follow the exact ground state potential energy surface, (\emph{ii}) the \emph{statistical} error introduces a noisy component in the forces leading eventually to instability of the dynamics. These two effects are investigated in Section~\ref{sec:results}.

\subsection{\label{ssc:Langevin} Generalized Langevin dynamics}

The second possibility we explore is using forces to perform finite temperature simulations.  
Langevin dynamics (LD) has been originally introduced in a MD context to simulate the diffusion of bodies immersed in bath of lighter particles~\cite{allen_computer_1987}.
In this framework, noisy and dissipative contributions are added to the Newton equation of motion described above.
However, LD can be employed also as a thermostat, to sample from a finite temperature canonical distribution
\begin{equation}
\label{e:boltz}
    \rho ( {\bR}, {\bv} ) \propto e^{ -\beta H( {\bR}, {\bv} )},
\end{equation}
at finite temperature $T = 1 /(k_B \beta)$~\cite{allen_computer_1987,bussi_accurate_2007} with $k_B$ being the Boltzmann constant.
We will show how the Langevin framework is particularly convenient when the forces are affected by statistical error bars, as in the present setting.
Here, one can exploit the freedom given by the fluctuation-dissipation theorem which sets the relation between the friction matrix and the power spectrum of the noise.
We notice that this possibility has been already put forward in the context of ab-initio MD~\cite{kuhne_efficient_2007,Attaccalite08}.
For instance, Sorella and coworkers~\cite{Attaccalite08,mazzola2014unexpectedly, luo2014ab} introduced this dynamics in  quantum Monte Carlo simulations, to cope with the fact that ionic forces are known only with a finite precision. 
This approach can be pursued both in the second order~\cite{Attaccalite08,luo2014ab} and in the first order Langevin dynamics cases~\cite{mazzola2012finite,mazzola2017accelerating,mazzola2018phase}.

In this paper we consider the second order Langevin dynamics as \emph{(i)} it smoothly connects to the Newtonian equation of motion in the limit of vanishing statistical sampling error, and \emph{(ii)} it has been shown that some dynamical properties can be rigorously computed from thermostatted trajectories in this case~\cite{rossi2018fine}.
The equations of motion read
\begin{eqnarray} 
\label{e:sld}
 \dot {\bv }   &=&  -   \bm \gamma(\bR) \cdot
  \bv   +   \bF(\bR)/{\bm m}  +\bm \eta  (t) \label{first}, \\
 \dot{  \bR }   &=&  \bv, \label{second} \\
 \left< \bm \eta(t) \right> &=& 0, \nonumber \\
 \left< \eta_i(t) \eta_j(t') \right> &=& \alpha_{ij}(\bR) \, \delta(t-t'), \nonumber
\end{eqnarray}
where $\bm \eta$ is a $3N$-dimensional vector representing the mass-rescaled Gaussian white noise force, with power spectrum given by $ {\bm \alpha}(\bm R)$ and $ \bm \gamma(\bm R)$ is a position dependent mass-rescaled friction matrix. 

The fluctuation-dissipation theorem 
\begin{equation}
\label{e2:fdt}
  {\bm \alpha}(\bR) =  2 T ~ \bm \gamma(\bR) ,
\end{equation}
dictates the relation between the friction matrix $\bm \gamma(\bR)$ and the noise $\bm \alpha(\bR)$, in order to sample the correct finite $T$ Boltzmann distribution of Eq.\eqref{e:boltz}~\cite{Becca2017, ceriotti_langevin_2009}.
In our context we can assume that the noise force $\bm \eta$ in Eq.~\eqref{e:sld} is  entirely due to the statistical noise intrinsic in the evaluation of the forces $\bF(\bR)$ with a finite numbers of shots (as defined above).
In this case, the matrix $\bm \alpha(\bR)$ is given by the covariance matrix of the forces, that can be computed at each time step according to Eq.~\eqref{eq:covmatgen} as
\begin{equation}\label{eq:covmat}
\alpha_{ij}(\bm R) = 
  \left< 
    \left(  F_i(\bm R) - \left< F_i(\bm R) \right> \right)
    \left(  F_j(\bm R) - \left< F_j(\bm R) \right> \right)
  \right>,
\end{equation}
and $\left< \cdots \right>$ indicates the statistical average over the measurements.
We notice that $\bm \alpha(\bR)$ is not constant during the dynamics as the variance of the forces is itself a stochastic quantity and varies during the simulations.

In this scheme, the value of friction is proportional to the noise fluctuations in the forces so that the corresponding ionic displacement is also anisotropically reduced, stabilizing the dynamics while sampling unbiasedly the canonical ensemble.

For the sake of demonstration we discretize Eq.~\eqref{e:sld} with the simplest Euler integrator, though much more sophisticated integration schemes exist~\cite{Attaccalite08,mazzola2014unexpectedly,luo2014ab}.
In addition, we simplify our method restricting ourselves only to the diagonal part of  $\bm{\alpha}(\bm R)$ and neglecting contribution from the off-diagonal elements.
We also mention that, in most general case, an external Gaussian distributed white noise can also be added in Eq.~\eqref{e:sld}, in addition to the intrinsic one present in the forces.
This can be useful to increase the friction, reaching an optimal value which minimizes the autocorrelation time of the simulations~\cite{Attaccalite08,mazzola2014unexpectedly,luo2014ab}.

\section{\label{sec:results}Results and Discussion}

To assess the quality of the force algorithm described above, we perform geometry optimization and MD studies (in the microcanonical and canonical ensembles) for two simple molecular systems:  H$_2$ and H$_3^+$ 
The size of these molecules allows for a systematic study of the different simulation conditions while keeping the computational costs relatively low.
In fact, while the quantum algorithms for electronic structure calculation show a favorable scaling comparing to the equivalent classical algorithms, their simulation with classical computers is far from efficient, especially when the classical calculation aims at reproducing the quantum variational approach, i.e., the VQE optimization as it would be implemented in a quantum computer.

To validate our approach, in the case of geometry optimization and microcanonical MD simulations we also provide a solution obtained using the matrix representation (MR) of the Hamiltonian and its direct diagonalization (Exact) to obtain the ground state energy and the corresponding eigenvector (wavefunction). 
On the other hand, to reproduce the conditions of a hardware calculation we also simulate the VQE algorithm using a realistic representation of the noise of one of the IBM Quantum processors, namely \textit{ibmq\_athens}. 
Insights about the dependence of the results on the level of noise are gained by repeating the simulations with a `noise model' that corresponds to a quantum device with improved gate fidelities and halved noise rates. 
The details about the noise models `full' and `half' are given in Table~\ref{tab:noise_model_parameters} of Appendix~\ref{sec:appendixB_noisesources}. 

For both systems, we use the STO-3G basis set. 
A HF/STO-3G calculation is performed to generate the molecular orbitals used to construct the Hamiltonian in second quantization. 
A total of 4 spin-orbitals (2 occupied and 2 virtual) were used for H$_2$ and 6 (2 occupied and 4 virtual) for H$_3^+$. 
The number of spin-orbitals is equal to the total number of qubits required for the simulations. 
To further reduce the computational costs, we applied qubit reductions schemes based on the symmetries of the qubit Hamiltonian. 
For the case of H$_2$, we apply the `two-qubit reduction' scheme~\cite{Bravyi2017} reducing the final number of qubits to 2. 
For the case of H$_3^+$, we use the restricted formalism for the expression of the HF orbitals (same spatial orbitals for both spin-up and spin-down molecular orbitals) reducing the number of qubits to 3 and we further apply the `tapering off qubits'~\cite{Bravyi2017} scheme to reduce to 2. 
A more elaborate description of our used techniques can be found in Appendix~\ref{sec:appendixA_ressources} and Ref.~\cite{Bravyi2017}.

\begin{figure}[ht]
\centering
\includegraphics[width=1.\linewidth]{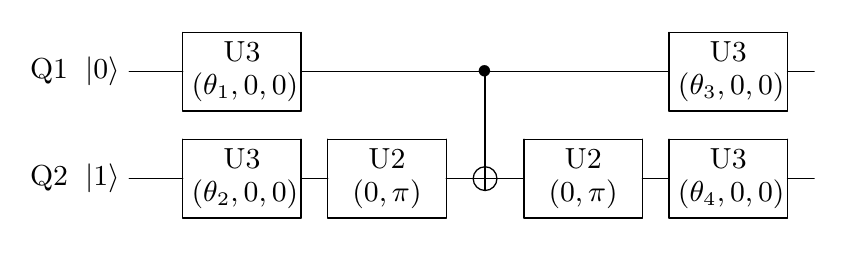}
\caption{{  
Quantum circuit corresponding to the RY Ansatz. 
The VQE variational parameters $\vec{\theta} = (\theta_1,\theta_2,\theta_3,\theta_4)$ control the rotations on the Bloch sphere with $U3(\theta, \phi, \lambda) = R_z(\phi-\pi/2) R_x(\pi/2) R_z(\pi-\theta) R_x(\pi/2) R_z(\lambda-\pi/2)$ and $U2(\phi, \lambda) = R_z(\phi+\pi/2)R_x(\frac{\pi}{2})R_z(\lambda-\pi/2)$.
In particular, $U3(\theta, 0, 0) = R_y(\theta)$ hence the name of the Ansatz. For further details refer to Appendix~\ref{sec:appendixC_RYAnsatz}.
}}
	\label{fig:ry_circuit}
\end{figure}

The wavefunction are expanded using the RY Ansatz~\cite{qiskit2019} with depth $1$, which amounts to a total of $4$ variational parameters both for H$_2$ and H$_3^+$ (see Fig.~\ref{fig:ry_circuit}). 
The classical optimization of the VQE parameters was performed using the COBYLA optimizer~\cite{cobyla} with default settings as defined in SciPy software library~\cite{virtanen2020scipy}.

All MD calculations are performed using the finite difference formula defined in Eq.~\eqref{eq:forces_HF}. The integration of the equations of motion is done using the Verlet algorithm (Eq.~\eqref{eq:verlet}) and a time step of $0.2$ fs for microcanonical MD and the Langevin algorithm (Eq.~\eqref{e:sld}) 

For the Lanczos algorithm (see Eq.~\eqref{eq:lanczos1}), we use $d_{PES}=-0.4$ for the evaluation of the potential energy and $d_{forces} = -0.1$ for the forces.
The noise models used in the simulations are summarized in Appendix~\ref{sec:appendixB_noisesources}, while the implemented measurement (readout) error mitigation scheme is described in the documentation of Qiskit~\cite{qiskit2019}.

In the case of H$_2$, we also performed a short ($20$ fs long) hardware microcanonical MD calculation using the IBM chip \textit{ibmq\_athens}. The characterization of this device are given in Appendix~\ref{sec:appendixB_noisesources}.

\subsection{Validation of the H-F forces calculation}

\begin{figure}[ht]
\centering
\includegraphics[width=1.\linewidth]{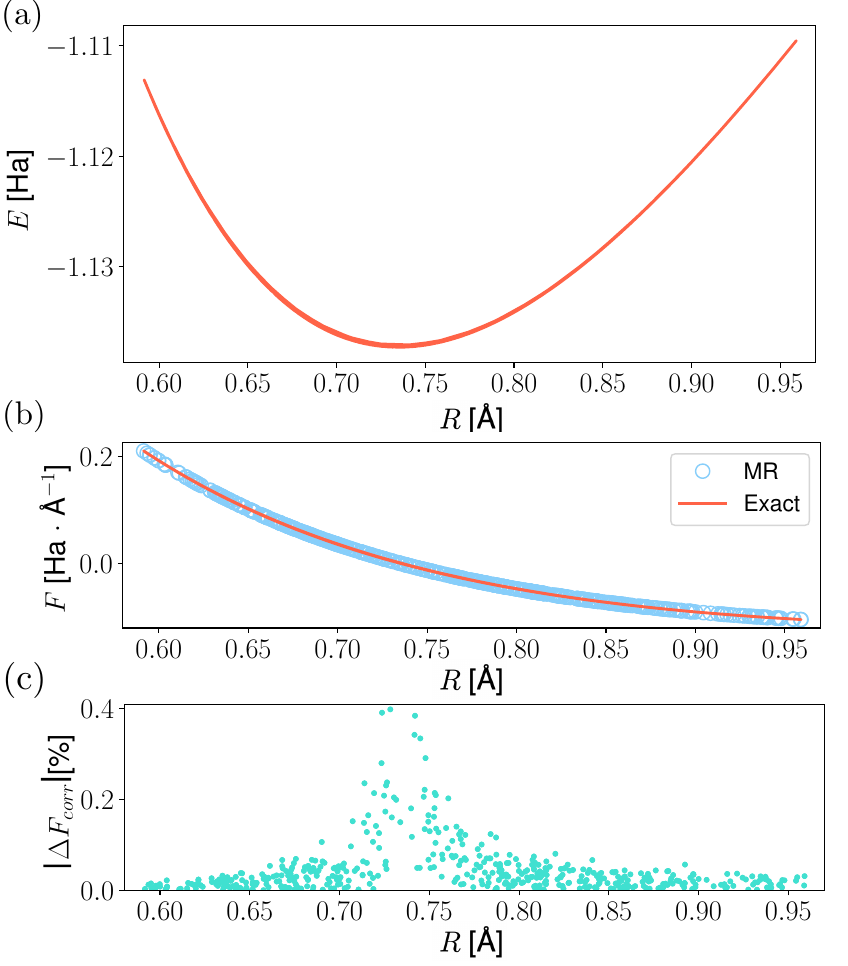}
\caption{{(a) Potential energy obtained by exact diagonalization of the H$_2$ Hamiltonian;
(b) Forces along the molecular axes calculated with finite differences on the curve in (a) (denoted by `Exact') and with the H-F approach (see Eq.~\eqref{eq:forces_HF}, denoted by `MR'). 
A displacement $\Delta R = 10^{-3}$\AA ~is used in both cases; 
(c) Percent error on the forces due to the omission of the Pulay component.
}}
	\label{Figure_micro_MD_H2_HFvalidation}
\end{figure}

To verify the quality of the H-F forces and estimate the error done by neglecting the last two components in Eq.~\eqref{eq:forces} (Pulay's forces), we performed an exact calculation of the forces for H$_2$ by means of the direct differentiation (using finite differences with a displacement of $10^{-3}$~\AA) of the potential energy surface obtained by exact diagonalization of the system Hamiltonian (Fig.~\ref{Figure_micro_MD_H2_HFvalidation}(a)).  
These noiseless forces are then compared to the H-F ones computed with Eq.~\eqref{eq:forces_HF} (Fig.~\ref{Figure_micro_MD_H2_HFvalidation}{(b)}), while the absolute percental error computed as $|\Delta \bm{F}_{corr}| = |\bm{F}_{exact} - \bm{F}_{MR}|/ |\bm{F}_{MR}|$, is reported in panel ({c}). 
The maximal deviation is observed for geometries close to the equilibrium bond distance and never exceeds $0.4\%$ of the total force.
Since the error due to statistical sampling as well as the hardware noise are much larger than this value, we can safely neglect the Pulay contributions for the remainder of this work.

\subsection{Geometry optimization of H$_2$ and H$_3^+$}

In this section, we use the most accurate force calculation setup, i.e., Eq.~\eqref{eq:forces_HF} with Lanczos noise mitigation, to evaluate the optimized geometry for H$_2$ and H$_3^+$ using a realistic model of the \textit{ibmq\_athens} device. 
The results are summarized in Table~\ref{tab:geomopt}. 
\begin{table}[ht]
\caption{{Geometry optimization results using the MR (reference), the VQE and the VQE with Lanczos (VQE-L) algorithms. 
In these two last cases we use 8'192 measurements for the evaluation of the energy and force components. 
The equilibrium bond distance of H$_2$ is given by $R_{eq}$.
The structure of H$_3^{+}$ (see Fig.~\ref{fig:schema_md}) is characterized by three parameters: (\textit{i}) the distance between atoms 1 and 2 ($R_{12}$),  
(\textit{ii}) the distance between atoms 1 and 3 ($R_{13}$),
and 
(\textit{iii}) the angle $\alpha_{213}$ formed between the bonds H$_1$-H$_2$, and H$_1$-H$_3$ ($\alpha_{213}$).
We use \AA \ for distances and degrees for the angles.
}}
  \begin{tabular}{p{0.04\textwidth} p{0.06\textwidth} p{0.1\textwidth} p{0.1\textwidth} p{0.12\textwidth}}
    \hline
    \hline
 & & MR & VQE  & VQE-L  \\ 
 \hline 
 H$_2$&    $R_{eq}$   & 0.735  &  0.742   & 0.733   \\
 \hline
H$_{3}^{+}$& $R_{12}$  & 0.985 & 1.006 &  0.990 \\
& $R_{13}$ & 0.985 & 0.999  &  0.990 \\
& $\alpha_{213}$ & 60.0  & 59.8  &  59.9 \\
    \hline
    \hline
  \end{tabular}
\label{tab:geomopt}
\end{table}

We observe that the forces evaluated with the standard VQE using the noise model of a real device (\textit{ibmq\_athens}) have errors in the order of $0.01$-$0.02$~\AA~ in the bond distances and $0.2^{
\circ}$ in the angle $\alpha_{213}$ (see Fig.~\ref{fig:schema_md}).
By using the Lanczos mitigation scheme, we improve significantly the quality of the forces and, consequently, the quality of the optimized structures (i.e. reducing the error on bond distances to $0.002$-$0.005$~\AA). 
As it will become evident in the study of the MD, even though the error on the energies are still not in the so-called chemical accuracy (1 kcal/mol or $1.6$ mHa), the forces computed directly using Eq.~\eqref{eq:forces_HF} show a great level of accuracy and can be used to compute optimized molecular geometries.

\subsection{Microcanonical dynamics of H$_2$ and H$_3^+$}\label{Results_microcan_H2_H3+}

In this section, we apply the forces derived in Eq.~\eqref{eq:forces_HF} for the calculation of microcanonical MD trajectories of H$_2$ and H$_3^+$. 

\begin{figure*}[ht]
\centering
\includegraphics[width=1\linewidth]{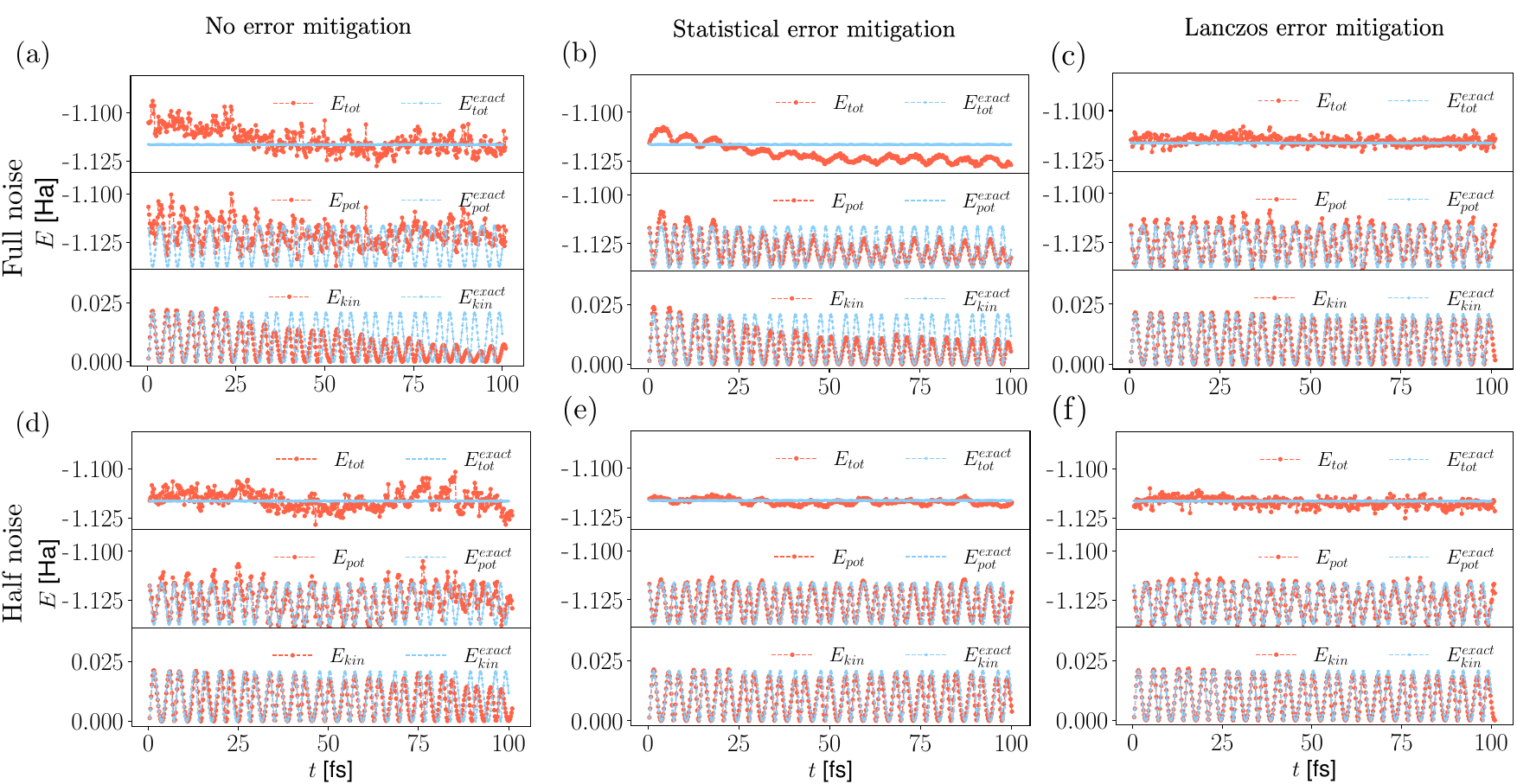}
\caption{{
Time series of the total, potential and kinetic energies for the dynamics of the H$_2$ molecule ($t = 100$ fs, $dt= 0.2$ fs) using the VQE algorithm with the realistic noise corresponding to the \textit{ibmq\_athens} device and: 
(a) 8'192 measurements for the evaluation of energies and forces,
(b) 81'920 measurements for the evaluation of energies and forces,
(c) the VQE-L error mitigation scheme with 8'192 measurements for the evaluation of energies and forces.
Panels (d), (e), (f) report the same simulations as in (a), (b), (c) with halved hardware noise level.
In all panels, the blue curves correspond to the reference `exact' dynamics obtained with the MR.
}}
	\label{fig:h2_all_in_one}
\end{figure*}

In the first case (H$_2$), we perform a systematic study of the effect of the different noise sources (hardware and statistical) on the quality of the dynamics. 
All calculations of this section are performed using a noise model corresponding to the \textit{ibmq\_athens} device that is summarized in Table~\ref{tab:noise_model_parameters}, in Appendix~\ref{sec:appendixB_noisesources}. 
To shed light on the potential future improvements associated to the development of future hardware, we also present results for the half of the current hardware noise.
In particular, we present the time series of kinetic, potential and total energies for three different simulations conditions: 
(\textit{i}) VQE with 8'192 measurements for the determination of the energy and force expectation values;
(\textit{ii}) VQE with 81'920 measurements for an improved accuracy of both energy and force estimations and corresponding decrease of the statistical noise;
(\textit{iii}) VQE with 8'192 measurements combined with the Lanczos noise mitigation algorithm with $d_{PES}=-0.4$ and $d_{forces}=-0.1$ (see Eq.~\eqref{eq:lanczos1}).
Note that the choice of 8'192 measurements is motivated by the limitation of current IBM quantum hardware, which cannot exceed this number.
In all cases, we will take the microcanonical MR trajectory as reference, exact dynamics of the system.
The microcanonical dynamics were performed using the Verlet algorithm described in Section~\ref{sec:Verlet} and using an initial bond length of 0.6~\AA.
The results of these three simulations are reported in Fig.~\ref{fig:h2_all_in_one}(a), Fig.~\ref{fig:h2_all_in_one}(b), and Fig.~\ref{fig:h2_all_in_one}(c), respectively.

In all cases, we observe that in all VQE \textit{noisy} simulations the accuracy of the kinetic energy is much higher than the one for the potential energy. 
Here and in the following we use the adjective \textit{noisy} to emphasize the use of noise model in the classical simulations of the VQE algorithm.
This is possible because the forces are computed independently from the energy using the expectation value elements in Eq.~\eqref{eq:forces_HF}.
As a result, we obtain reliable dynamics even when the overall error in the potential energy is higher than what would be required for numerical differentiation. 
In the case of bare VQE simulation (Figs.~\ref{fig:h2_all_in_one}(a),~\ref{fig:h2_all_in_one}(d)), the full noise dynamics reproduces quite accurately the vibrational frequency of the bond oscillation even though an important damping is also observed, which is not compensated by an equal increase in the potential energy.
As a consequence the total energy is not conserved (Fig.~\ref{fig:h2_all_in_one}(a)). 
However, when further prolonging the simulation over the 100 fs shown here (see Fig.~\ref{fig:h2_verlet_1200fs} in Appendix~\ref{sec:appendixE_longverlet}) we observe an oscillatory behaviour of total energy over a time scale of about 1 ps. 
We can therefore interpret the noise as an external bath that modulates the total energy of the H$_2$ subsystem. 
We will fully exploit this property in the Langevin dynamics reported in the next section (Section~\ref{Results_Langevin_H2}).
When halving the level of noise (Fig.~\ref{fig:h2_all_in_one}(d)), we observe an important improvement of the total energy conservation and a more accurate oscillation frequency compared to the reference, noiseless, MR calculation. 
This indicates that the hardware noise is currently among the most important limitations for microcanonical dynamics, even though with the superconducting qubit technology we are not far from reaching the accuracy necessary for simulations in the picosecond timescale (see also the results of the hardware MD calculation of H$_2$ in Fig.~\ref{Fig_H2_hardware} of Section~\ref{Results_hardware_H2}).

In Figs.~\ref{fig:h2_all_in_one}(b),~\ref{fig:h2_all_in_one}(e), we present the case in which we increase the sampling statistics for the calculation of the energy and force expectation values from 8'192 to 81'920. 
As expected, we observe a general decrease of the overall scattering of the energy values due to the improved statistics.
However, the same issue with the conservation of the total energy observed in Figs.~\ref{fig:h2_all_in_one}(b),~\ref{fig:h2_all_in_one}(e) persist in both case, full and halved noise levels, even though less severe in the last case.
Once more, this corroborates the hypothesis that the hardware noise (and not the statistic sampling) is the main source for the non-conservation of the total energy.

Finally, in Figs.~\ref{fig:h2_all_in_one}(c),~\ref{fig:h2_all_in_one}(f) we report the simulation using the Lanczos noise mitigation scheme (VQE-L). For both levels of noise, the reproduction of the reference MR dynamics is extremely good, despite the use of the minimal sampling statistics, i.e., 8'192 measurements for each expectation value. The conservation of total energy is maintained over the total simulation length and the bond frequency is captured very closely. 

\begin{figure}[ht]
\centering
\includegraphics[width=1.\linewidth]{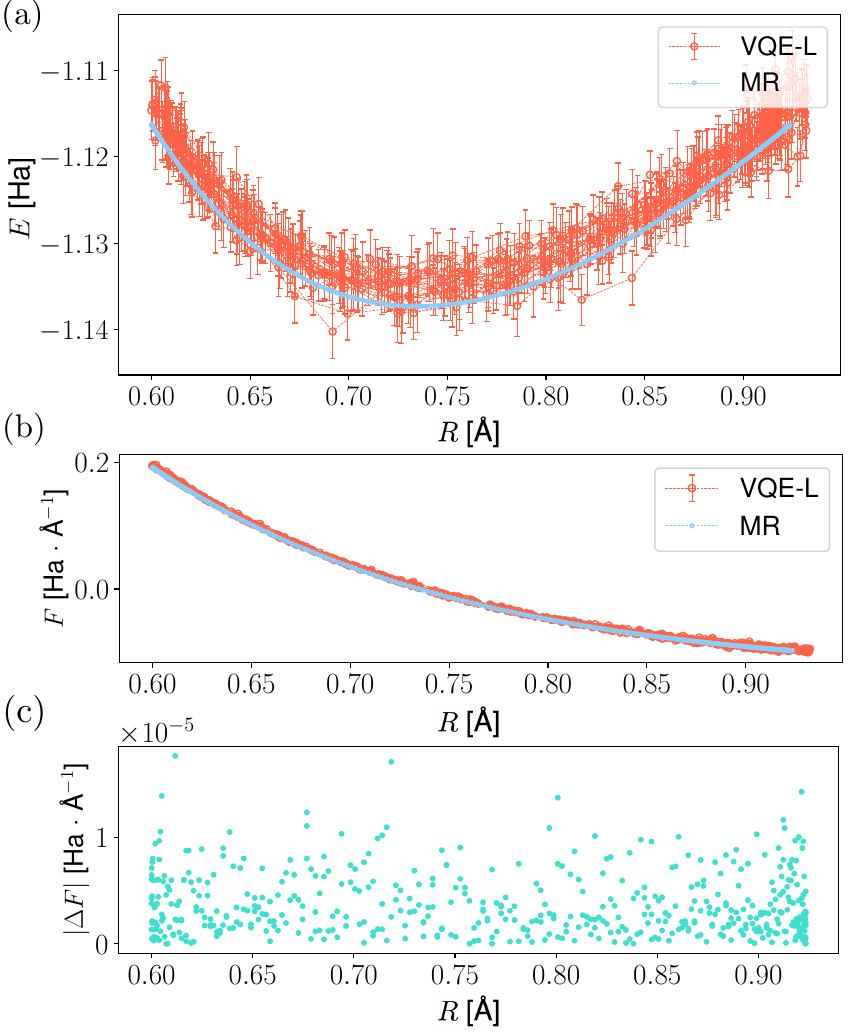}
\caption{{
(a) Sampling of the potential energy obtained with the 100 fs VQE-L dynamics of H$_2$ shown in Figure~\ref{fig:h2_all_in_one}(e). The blue line correspond to the MR result. 
(b) Corresponding forces along the molecular axes acting on one of the hydrogen atom as a function of the bond distance. 
(c) Norm of the difference between the forces computed with VQE-L and MR.
MR forces were recomputed at the same geometries sampled along the VQE-L trajectory.
}}
	\label{Figure_micro_MD_H2_Lanczos_2}
\end{figure}

In Figure~\ref{Figure_micro_MD_H2_Lanczos_2}, we give a overview on the performance of the VQE-L algorithm for microcanonical MD by reporting each energy value of the structures sampled along the trajectory 
in Fig.~\ref{fig:h2_all_in_one}(c) as a function of the corresponding bond length. 
Once more we notice that even though the error in the potential energy is of the order of several mHa, using Eq.~\eqref{eq:forces_HF} we can obtain fairly reliable forces (see Fig.~\ref{Figure_micro_MD_H2_Lanczos_2}(b)), which can be used for accurate geometry optimization and MD simulations.
The components of the forces acting on one of the hydrogen atoms along the molecular axes are shown in Fig.~\ref{Figure_micro_MD_H2_Lanczos_2}(b) together with the reference H-F forces obtained with the MR approach. 
Finally, in Fig.~\ref{Figure_micro_MD_H2_Lanczos_2}(c) we report the difference between the H-F forces computed with VQE-L and the ones obtained with the reference MR approach.  
We notice that, for this particular example, the magnitude of the error on the forces is approximately equally spread over the entire bond range with a magnitude smaller than $2.0 \times 10^{-5}$ Ha $\cdot$ \AA$^{-1}$, which is already a fairly good accuracy considering the level of noise of the quantum processors. 

\subsection{Microcanonical dynamics of H$_2$ on a quantum computer}\label{Results_hardware_H2}

As a final demonstration of the accuracy of the proposed algorithm, we perform constant energy dynamics of H$_2$ on the IBM quantum computer \textit{ibmq\_athens}. 
The same approach could be also applied to the Langevin dynamics; however, due to the long equilibration times this would require a large number of steps that we cannot yet afford. 
The characteristic features of this machine at the time of execution (13 July 2020) are summarized in Table~\ref{tab:noise_model_parameters}, Appendix~\ref{sec:appendixB_noisesources}.

\begin{figure}[ht]
\centering
\includegraphics[width=1\linewidth]{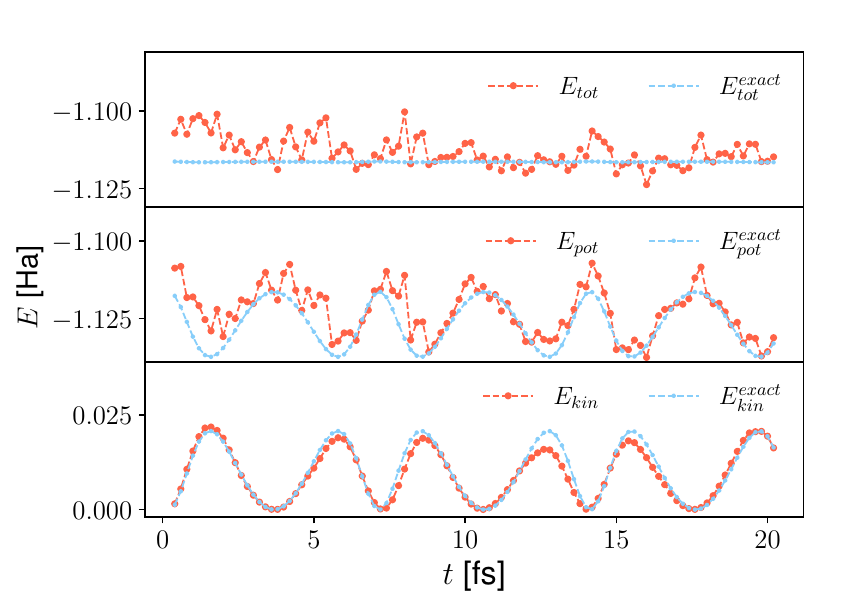}
\caption{{Time series of the total, potential and kinetic energies for the dynamics of the H$_2$ molecule on the IBM quantum computer \textit{ibmq\_athens}. 
Calculations were done using the bare VQE algorithm (without Lanczos noise mitigation) with 8'192 measurements per expectation value.  
The equation of motion was solved with the velocity-free Verlet algorithm and a time step of $0.2$ fs. 
}}
	\label{Fig_H2_hardware}
\end{figure}

The dynamics on hardware obtained with the bare VQE algorithm (without Lanczos noise mitigation) and the velocity-free Verlet integration scheme with a time step of 0.2  fs 
confirm the quality of the noise models used in the simulations reported in Section~\ref{Results_microcan_H2_H3+} (see Fig.~\ref{Fig_H2_hardware}).
Once more, we observe that despite the quite large deviation for the calculation of the potential energy (of about 10-15 mHa) using the canonical 8'192 measurements for each expectation value the accuracy of the kinetic energy is an order of magnitude higher. 
This is due to the stability of the force calculation using the expectation values introduced in Eq.~\eqref{eq:forces_HF}, which bypasses the derivative of the noisy potential energy surface.

The stability of this microcanonical dynamics scheme is such that we even do not need to apply the Lanczos noise mitigation scheme, which will require the evaluation of additional expectation values with an impact on the overall machine time. 

\subsection{Microcanonical dynamics of H$_3^+$}

\begin{figure}[ht]
\centering
\includegraphics[width=1.\linewidth]{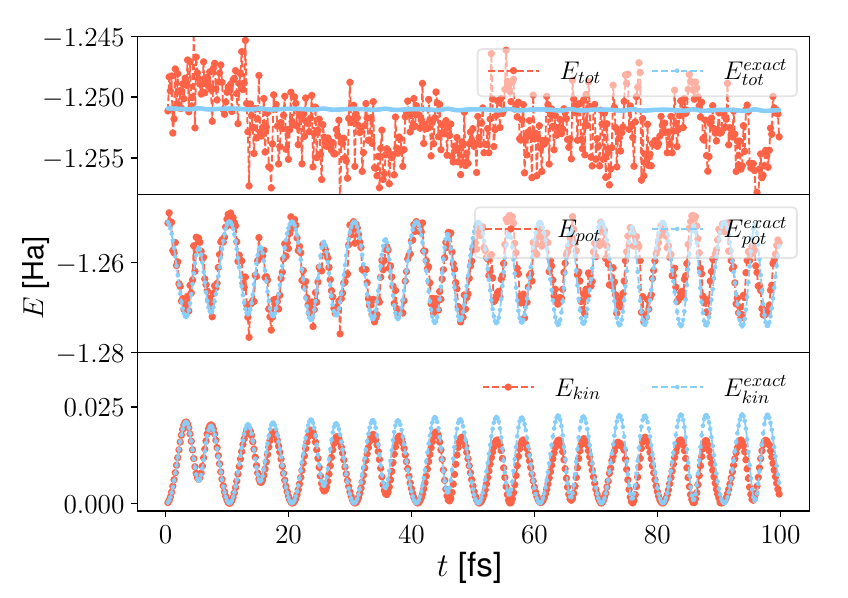}
\caption{{Evolution of the total, potential and kinetic energies during the dynamics of the H$_{3}^{+}$ molecule ($t = 100$ fs, $dt= 0.2$ fs). 
Using the VQE algorithm with realistic noise of $ibmq\_athens$ chip and 8'192 shots, we demonstrate the simulation with Lanczos method (fixed $d$ method, $d_{PES}=-0.4$ and $d_{Forces}=-0.1$). Details for the calculation are described in Section~\ref{sec:results}.}}
	\label{fig:H3_energies}
\end{figure}

The H$_3^+$ molecule is an interesting system for the validation of our MD algorithm. While conserving the same number of electrons as in H$_2$, the size of the configuration space increases to 9 (or 6 if we remove the translational degrees of freedom) introducing additional dynamical degrees of freedom, such as bond vibrations. 
Building from what we learned in the previous section, we will restrict the study of the microcanonical dynamics of H$_3^+$ to the use of the VQE-L algorithm with the velocity-free Verlet for the integration of the equations of motion.
As initial conditions we take an off-equilibrium geometry characterized by the internal parameters $d_{12}= 1.245$ ~\AA,  $d_{13}= 1.245$~\AA, and $\alpha_{213}=48.5^\circ$ (see Table~\ref{tab:geomopt} for the corresponding equilibrium values) and zero velocities for all atoms.
The time series for the kinetic, potential and total energies obtained over 100 fs of NVE dynamics are given in Fig.~\ref{fig:H3_energies}.
As for the case of H$_3^+$, we observe a fairly good energy conservation with a drift of about $5$ mHa over 100 fs of dynamics. 
Also important is the accuracy with which the dynamics can capture the non-trivial molecular oscillations, as illustrated by the evolution of the kinetic energy (bottom panel in Fig.~\ref{fig:H3_energies}) and the agreement with the reference MR calculations.

\subsection{Langevin dynamics of H$_2$ }\label{Results_Langevin_H2}

\begin{figure*}[ht]
\centering
\includegraphics[width=1\linewidth]{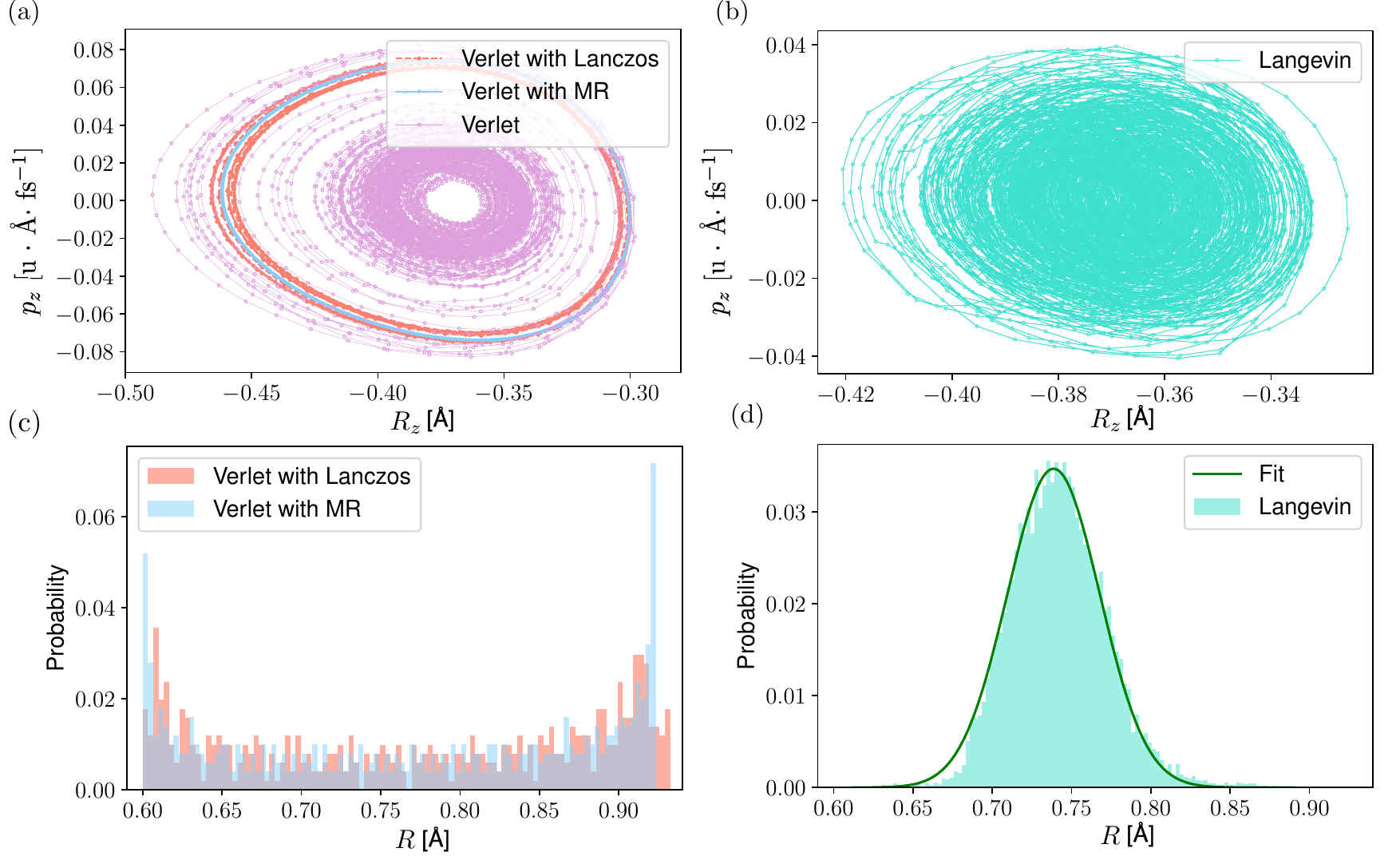}
\caption{
{Color) Phase space orbits corresponding to one of the hydrogen atoms in H$_2$: (a) Microcanonical dynamics with MR, VQE, VQE-L; (b) Constant temperature Langevin dynamics at an average temperature $T = 423$ K.
Distribution of the bond length from the simulations with: (c) Verlet ($R_{eq} = 0.6$ \AA, 8'192 shots) and (d) Langevin dynamics ($R_{eq} = 0.73$ \AA, 1'024 shots).
In the microcanonical case, the noisy and noiseless simulations are denoted with `Lanczos' and `MR', respectively. The parameters for the Lanczos mitigation are described in Section~\ref{sec:forces_methods}.
The Langevin canonical bond distribution (in panel (d)) is fitted to a Gaussian function $\rho ({R}) = a \, e^{ -\beta (\frac{1}{2}k(R-b)^{2})}$, with parameters $a = 3.397 \cdot 10^{-2}$, $b = 0.7387$ \AA~and $\beta = 680.6$ Ha$^{-1}$. 
The value of $k=1.681$ Ha $\cdot$ \AA$^{-2}$~was obtained from a separate fit of the PES obtained with the MR approach.
}}
	\label{fig:lang_combined}
\end{figure*}

In this Section, we demonstrate the Langevin dynamics driven by the statistical noise on the evaluation of the forces (see Sections~\ref{ss:gs} and~\ref{ssc:Langevin}) on H$_2$.
To this end we observe that the Eq.~\eqref{e:sld} defines a stochastic process having as fixed point the unique equilibrium distribution Eq.~\eqref{e:boltz}.
In Figure~\ref{fig:lang_combined}(d), we plot the distribution of the bond lengths $R$ sampled via a Langevin dynamics of total length $t=2.8$ ps using 1'024 shots and $dt=0.2$ fs, resulting in a measured temperature of $T_{kin}= 420 \pm 100$ K (from the expectation values of the kinetic energy).
Statistical error bars on the measured temperature are obtained with the standard binning technique~\cite{Becca2017} and can be decreased by running longer simulations. 
By contrast the number of shots used to compute the expectation values controls the friction value, hence the autocorrelation time of the sampling.

The normalized distribution $\rho(R)$ that we obtain, with $R$ being the bond length, can be fitted by the canonical distribution  Eq.~\eqref{e:boltz} (marginalized over the relative momentum $p_z$), $\exp{(-\beta U(R))}$, where $U(R)$ is a quadratic fit of the MR potential energy surface of H$_2$, and $T = 1 / (k_B\beta) = 423 \pm 30$ K is a fitted temperature (error estimated as the error due to 3 standard deviations on the fitting parameters). 
The agreement between $T$ and $T_{kin}$ shows that a finite temperature distribution can be achieved by means of the Langevin equation.

On the other hand, it is qualitatively evident that the distribution of bond lenghts produced by the noisy Verlet dynamics (see Fig.~\ref{fig:lang_combined}(c)) is not in agreement neither with a Boltzmann distribution which is peaked around $R_{eq}$, nor with the microcanonical one which is sharply peaked on the turning points of the trajectory.

The difference in the sampling between the Langevin and Verlet integrator is also shown in Fig.~\ref{fig:lang_combined}.
Panel ({a}) shows the phase space orbits of one of the two hydrogen atoms of the H$_2$ molecule for different microcanonical simulations. 
The `exact' dynamics obtained in the MR representation of the quantum circuit (in blue) shows the well-defined orbit typical of a constant energy MD. 
The trajectory obtained with the VQE-L algorithm (in orange) is covering a thin annulus in the phase space that closely follows the exact orbit. 
This behavior can be associated to the small total energy fluctuations observed in Fig.~\ref{fig:h2_all_in_one}(c), which originate from the hardware noise. 
By removing the noise mitigation (violet curve), the microcanonical dynamics obtained with the Verlet algorithm under the influence of the hardware and statistical noises produces a dynamics that resembles a constant temperature MD leading to the sampling of thick annulus in phase space. 
Since in the first 100 fs of the bare VQE dynamics shows a drop of the total energy (Fig.~\ref{fig:h2_all_in_one}(a)), the orbit is mainly sampling the low energy region of the phase space.
As expected, all microcanonical MD sample a bimodal distribution of the molecular bond length with two maxima at the turning points (Fig.~\ref{fig:lang_combined}(c)).
Finally, in panel ({b}), we report the same phase space trajectory for the canonical Langevin dynamics of H$_2$. 
Due to the different choice of the initial bond length, the trajectory is sampling a different portion of the phase space while the coupling to the noise is keeping the system at an average temperature of about 423 K. 
Most importantly, the Langevin dynamics reproduces the correct Gaussian-like canonical distribution of the bond length (Fig.~\ref{fig:lang_combined}(d)).

\section{\label{sec:conclusion}Conclusion}

In this work, we implemented a quantum algorithm for the calculation of accurate forces within the VQE framework. 
To this end, we used the ground state wavefunction to directly measure the expectation values of the force operator associated to each atom of the system. 
In this way, we obtain high quality Hellmann-Feynman forces, which show larger robustness against hardware and readout noises than the energies itself.
We demonstrated that the quality of the forces is not affected by the omission of the components arising from the direct derivative of the ground state wavefunction (i.e., the Pulay components). 

In addition, by applying the Lanczos mitigation scheme on the forces evaluation we are able to achieve an accuracy that allows for the calculation of optimized molecular geometries with errors of about 0.005~\AA~for the bond lengths and of $0.1^\circ$ for the angles in two test cases: the H$_2$ and H$_3^+$ molecules.
Using the same algorithm for the force calculation, we have also performed molecular dynamics simulation of H$_2$ and H$_3^{+}$ in the constant energy (microcanonical) ensemble as well as in the constant temperature (canonical) ensemble using the Langevin approach. 
All calculations were performed using a realistic description of the hardware and readout errors corresponding to state-of-the-art IBM quantum computers.  
Despite the sizable errors in the evaluation of the potential energy, the direct calculation of the forces enables the calculation of accurate trajectories as demonstrated by the accuracy of the kinetic energy profiles, which is in good agreement with the reference calculations. Also in this case, the use of the Lanczos mitigation scheme in VQE (VQE-L) gives rise to a significant improvement of the dynamics, as can be seen from the conservation of the total energy of the system. 
We also stress the fact that by increasing the number of sampling points (readout measurements) or halving the level of noise we obtain a noticeable improvement on the quality of the simulations. In particular, the VQE-L approach with full hardware noise provides forces with an accuracy of about $10^{-5}$ Ha $\cdot$ \AA$^{-1}$. 
Motivated by the promising results obtained with the simulations, we also performed 20 fs of dynamics of H$_2$ on the \textit{ibmq\_athens} device obtaining stable and accurate description of the molecular vibration. 

Finally, we investigated the possibility to take advantage of the statistical noise inherent to the measurement process of the expectation values of the forces to perform constant temperature MD using the Langevin algorithm. 
In particular, we showed that by a suitable choice of measurement noise level through the number of shots we can control the autocorrelation time of the simulations and accurately sample the Boltzmann distribution for the H$_2$ molecule at a given temperature. However, further investigations are necessary to improve the coupling to the reservoir (i.e. computation of off-diagonal elements in the force covariance matrix) and the corresponding selection of an optimal friction through a more refined integration scheme.

We can therefore conclude that, despite the sizable error in the evaluation of the ground state energy of molecular system, the quality of the forces evaluated using the VQE algorithm, including error mitigation schemes such as the proposed Lanczos approach, enables the calculation of accurate optimized geometries as well as stable constant energy and constant temperature MD trajectories, opening up new avenues for the use of quantum computers in molecular chemistry and physics.

\section{\label{sec:acknowledgements}Acknowledgements}

The authors acknowledge the financial support from the Swiss National Science Foundation (SNF) through the grant No. 200021-179312.
We also thank Prof. J\"urg Hutter for insightful discussions.
IBM, the IBM logo, and ibm.com are trademarks of International Business Machines Corp., registered in many jurisdictions worldwide. Other product and service names might be trademarks of IBM or other companies. The current list of IBM trademarks is available at {\url{https://www.ibm.com/legal/copytrade}}

\appendix

\renewcommand{\theequation}{\thesection\arabic{equation}}
\setcounter{equation}{0}

\section{One/two-electron integrals}\label{sec:appendixD_integrals}

The prefactors of the one-/two-body excitations in Eq.~\eqref{eq:H} are given respectively by one-/two-electron integrals in physics notation and on molecular orbital basis, written as
\begin{equation}\label{eq:1eint}
 h_{rs}(\bm{R}) = \int d \bm{r}_1 \, \phi_r^*(\bm{r}_1) \left(-\frac{1}{2} \nabla^2_{\bm{r}_1} - \sum^M_{I=1} \frac{Z_I}{R_{1I}}\right) \phi_s(\bm{r}_1), 
\end{equation}
and likewise for the two-electron terms, given by
\begin{equation}\label{eq:2eint}
g_{rstu}(\bm{R}) = \int d \bm{r}_1 d\bm{r}_2 \, \phi_r^*(\bm{r}_1) \phi_s^*(\bm{r}_2) \frac{1}{r_{12}} \phi_u(\bm{r}_1) \phi_t(\bm{r}_2),
\end{equation}
where $Z_I$ is the nuclear charge of atom $I$, $R_{1I}=|\bm{R}_I-\bm{r}_1|$ distance between the nuclei $I$ and electron $1$, and the distance between a pair of electrons $r_{12}=|\bm{r}_1-\bm{r}_2|$ where $\bm{r_1}$, $\bm{r_2}$ denote their positions.
In this work, the computation of elements defined Eqs.~\eqref{eq:1eint},~\eqref{eq:2eint} is performed using HF method in PySCF software~\cite{Sun2018}.

\section{Pauli matrices}\label{sec:appendixF_paulimat}

We use the following notation for the Pauli operators
\begin{align}
      \label{eq:pauli_spin_operators}
    \hat{I} \equiv \mathbb{I} = \left( \begin{array}{cc}
        1 & 0 \\
        0 & 1
    \end{array}
    \right),~~
    \hat{X} \equiv \hat{\sigma}^x = \left( \begin{array}{cc}
        0 & 1 \\
        1 & 0
    \end{array}
    \right),\\
    \hat{Y} \equiv \hat{\sigma}^y = \left( \begin{array}{cc}
        0 & -i \\
        i & 0
    \end{array}
    \right),~~ \notag
    \hat{Z} \equiv \hat{\sigma}^z = \left( \begin{array}{cc}
        0 & 1 \\
        1 & 0
    \end{array}
    \right).  
\end{align}

\section{Analysis of the noise sources}\label{sec:appendixB_noisesources}

In Table~\ref{tab:noise_model_parameters}, we provide the noise model used for the simulations in this work. 
We give the necessary information (T1 , T2, qubit frequencies, readout errors, error rates for single qubit and two-qubit gates per qubit) to be able to reconstruct our noise model using Qiskit~\cite{qiskit2019}. 

\begin{table*}[ht]
\caption{
Noise model parameters for \textit{ibmq\_athens}.
Data for calibration date of 07/13/2020 06:13:48 GMT+0200 (CET). 
In Fig.~\ref{fig:h2_all_in_one}, the noise model corresponding to this data is denoted by `Full noise'. 
`Half noise' denotes the same noise model but with halved error rates. 
The notation `cx0\_1' denotes a CNOT gate between qubits 0 (control) and 1 (target).}
\begin{tabular}{cccccccc}
\hline \hline
Qubit & T1 ($\mu s$)            & T2 ($\mu s$)            & Freq. (GHz)   & Readout error         & Single-qubit U2 error rate & CNOT error rate  \\
\hline
Q0    & 64.915  & 104.232 & 5.176 & 1.000e-2 & 2.586e-4 & cx0\_1: 7.982e-3  \\
Q1    & 62.179 & 73.999  & 5.267 & 2.000e-2 & 3.186e-4       & cx1\_0: 7.982e-3, cx1\_2: 8.136e-3 \\
Q2    & 83.292  & 100.716 & 5.052 & 2.333e-2 & 3.440e-4       & cx2\_1: 8.136e-3, cx2\_3: 7.404e-3 \\
Q3    & 104.360 & 23.284 & 4.856 & 1.667e-2 & 2.633e-4       & cx3\_2: 7.404e-3, cx3\_4: 1.331e-2 \\
Q4    & 85.217   & 87.416  & 5.117 & 1.000e-2 & 2.970e-4       & cx4\_3: 1.331e-2 \\
\hline \hline
\end{tabular}
\label{tab:noise_model_parameters}
\end{table*}

The error sources considered in our simulations are the depolarization, thermalization and readout errors.
Next we present a brief summary of how they are modelled in Qiskit: 
\begin{itemize}
    \item[(i)] The \textit{depolarization error} consists of driving the noiseless density matrix $\rho = |\Psi\rangle \langle \Psi |$, with $|\Psi\rangle$ being a pure state, to the general uncorrelated density matrix $\mathbf{1} / 2^{N_q}$  as
\begin{equation}
\rho_{d}=\gamma_{1} \text{Tr}[\rho]  \mathbf{1} / 2^{N_q} + \left(1-\gamma_{1}\right) \rho,
\end{equation}
with $N_q$ being the number of qubits and $\gamma_{1}$ representing the decay to the uncorrelated state. The decay is estimated using gate fidelities given in Table~\ref{tab:noise_model_parameters}.
\item[(ii)] The \textit{thermalization error} 
(e.g. general amplitude dampening and phase flip error) of a qubit can be written as decay towards the Fermi-Dirac distribution of ground and excited states based on their energy difference $\omega$
\begin{equation}
\rho_{t}= p|0\rangle\langle 0|+(1-p)| 1\rangle\langle 1|,
\end{equation}
with $p=(e^{\frac{-\omega}{k_{b} T}}+1)^{-1}$, $T$ being the temperature and $k_B$, the Boltzmann constant.
\item[(iii)] The \textit{readout error} is classically modelled by calibrating the so-called measurement error matrix that assigns to any $N_q$-qubit computational basis state $| i \rangle$ (i.e. correct state that should be obtained) a probability to readout all the states $| j \rangle$ (i.e. the states that are actually obtained due to noise), or concisely $\mathcal{P}(i|j)$ where $i,j$ are $N_q$-qubit bit-string.
For instance, in an ideal noiseless situation, this matrix would be characterized by its matrix elements $\mathcal{P}(i|j) = 1$ for $i = j $ and $\mathcal{P}(i|j) = 0$ for  $i \neq j $.
\end{itemize}

\begin{figure*}[ht]
\centering
\includegraphics[width=1.\linewidth]{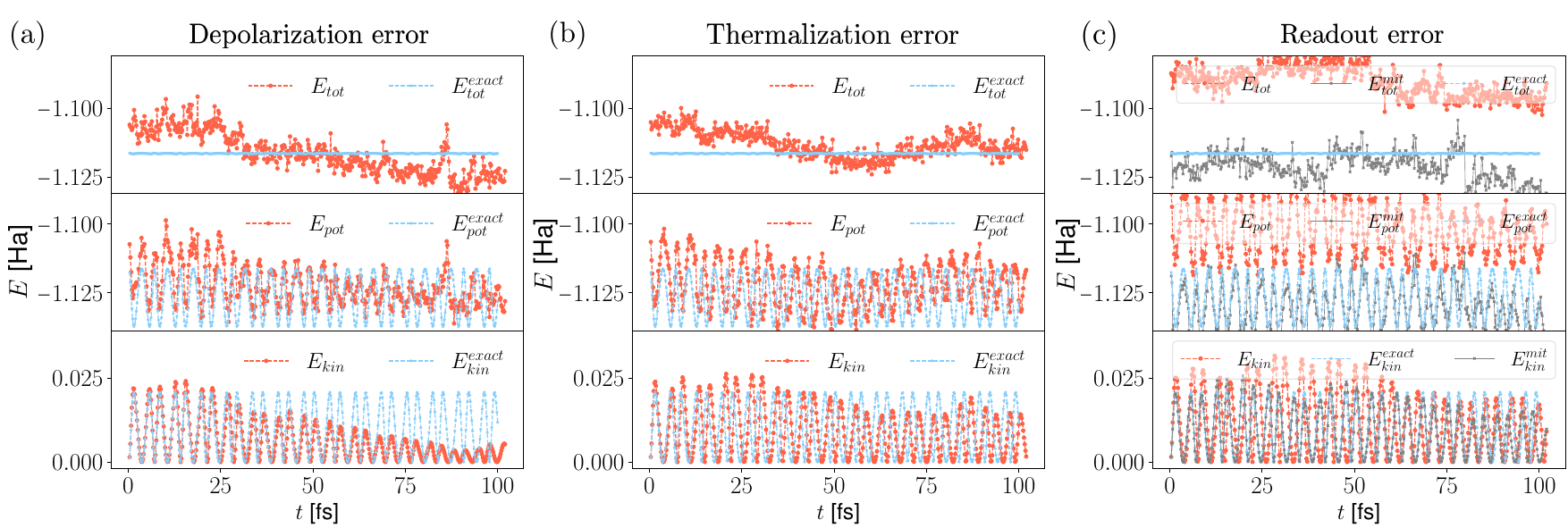}
\caption{{Evolution of the total, potential and kinetic energies during the dynamics of the H$_{2}$ molecule ($t = 100$ fs, $dt= 0.2$ fs). Using the VQE algorithm with realistic noise of $ibmq\_athens$ chip and 8'192 shots, we demonstrate the effects of individual components of the noise model, namely, (a) depolarization, (b) thermalization and (c) readout errors. 
In all panels, the blue curves correspond to the reference `exact' dynamics obtained with the MR.
In grey, the result of the simulation with readout error and measurement mitigation active.
Details of the calculation are described in Section~\ref{sec:results} and discussions are presented in Appendix~\ref{sec:appendixB_noisesources}.}}
	\label{fig:h2_noise_study}
\end{figure*}

To identify the major causes for the decay of kinetic energy in the H$_2$ simulations with noise (see Fig.~\ref{fig:h2_all_in_one}(a)),
we investigate the individual contributions of the different noise sources.
To this end, we perform a new series of MD simulations in which we activate a single noise source at a time (modelled from the \textit{ibmq\_athens} chip) reporting the results in Figure~\ref{fig:h2_noise_study}.
The settings remain the same as for the microcanonical simulations of H$_2$ reported in Section~\ref{sec:results}. The reference values obtained with the MR approach are in blue.
Figure~\ref{fig:h2_noise_study} shows that the depolarization error contributes the most to the loss of kinetic energy out of all other error sources.
In Fig.~\ref{fig:h2_noise_study}(a), we observe that the depolarization error contributes the most to the loss of kinetic energy.
The same is true for the total energy for which no recovery is observed over the entire simulation length (100 fs).
The total energy loss amounts to about $10$ mHa every 50 fs.
The thermalization error (Fig.~\ref{fig:h2_noise_study}(b)) has a significantly smaller impact than the depolarization error on the decay of kinetic energy, while the potential energies follows quite closely the reference profile.
For the case of the readout error, the potential energy presents a significant shift in order of $20$ mHa (Fig.~\ref{fig:h2_noise_study}(c)).
The use of measurement error mitigation corrects this issue (Fig.~\ref{fig:h2_noise_study}(c), grey curve) and provides PES and forces that are significantly improved, in better agreement with the exact reference.

\section{Selection of the Lanczos parameter}\label{sec:appendixE_lancz}

\begin{figure}[ht]
\centering
\includegraphics[width=0.9\linewidth]{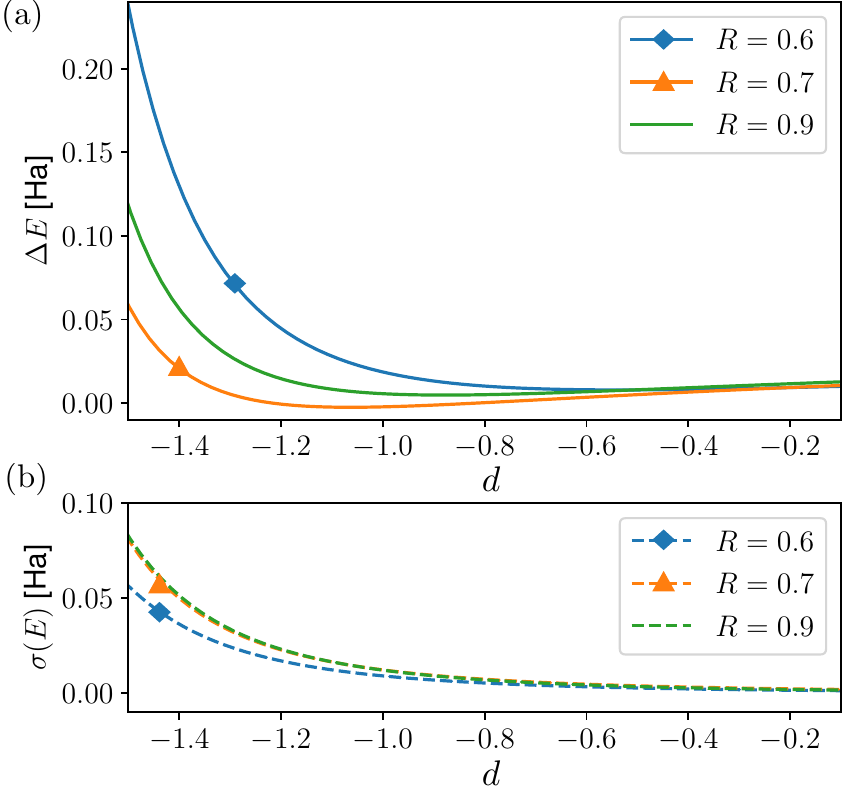}
\caption{{(a) Error $\Delta E$ and (b) standard deviation $\sigma (E)$ of the energy expectation value computed with 8'192 shots for H$_2$ molecule with bond lengths $R = 0.6, 0.7,0.9$~\AA~using the Lanczos method with different values of parameter $d$.
}}
	\label{H2_lanczos_d}
\end{figure}

In this section, we provide information about the selection of the parameter $d$ in the Lanczos approach. 
To this end, we perform a scan in the parameter $d$ for the system energy using 8'192 measurement (shots). 
For the case of the H$_2$ molecule, the test is repeated for three different bond distances: $0.6$~\AA, $0.7$~\AA~and $0.9$~\AA~(see Fig.~\ref{H2_lanczos_d}). 
All other simulation parameters are specified in Section~\ref{sec:results}.
We observe that from $d>-0.75$ the error on the energy $\Delta E$ (namely, the difference between the actual measurement on the hardware and the exact energy value obtained from the diagonalisation of Hamiltonian) start to increase linearly. 
Smaller values of $d$ (i.e., $d<-0.75$) leads to significant increase in the error on the energy at all bond distances together with an increase of the standard deviation associated to it, $\sigma(E)$. 
In order to guarantee accurate simulations for the ensemble or structures sampled during MD, we decided to take an intermediate value for $d$. 
Based on Fig.~\ref{H2_lanczos_d}, we selected a value of $d_{PES} = -0.4$, which produces accurate energy values at all geometries.
Note that this value is specific for the calculation of the energy expectation values for potential energy surfaces (from which the subscript ${PES}$ in $d_{PES}$). A similar approach is needed to select the corresponding optimal $d$ value for the forces calculations. 
The same procedure is then also applied to the case of the H$_3^{+}$ molecule.

\section{Reduction of quantum resources}\label{sec:appendixA_ressources}

For the case of H$_2$, to reduce the required quantum resources, the fermionic operators $\hat{a}^{\dagger}_r$, $\hat{a}_i$ in the molecular Hamiltonian $\hat{H}(\bm{R})$ are mapped to Pauli strings $P_{\lambda}$ (see Eq.~\eqref{eq:paulis_decomposition}) using the parity transformation~\cite{Bravyi2017}. 
This fermion-to-qubit mapping allows us to exploit the particle-number symmetry, $[\hat{H}(\bm{R}),\hat{N}_{\uparrow}]=[\hat{H}(\bm{R}),\hat{N}_{\downarrow}]=0 $ with $\hat{N}_{\sigma}$ being the number operator for electrons of spin $\sigma \in \{ \uparrow, \downarrow\}$.
This technique is a part of the `tapering off' procedure \cite{Bravyi2017} which permits the elimination of a qubit for each symmetry found in the Pauli string representation of the Hamiltonian.
Hence, as the commutators described right above correspond to two symmetries, two qubits can be eliminated from the simulation without modifying the spectrum of the Hamiltonian.
For the detailed explanation, we refer to the original work of Bravyi et al.~\cite{Bravyi2017} and, in context of quantum chemistry, see Appendix F in Ref.~\cite{sokolov2020quantum}.

For H$_{3}^{+}$ simulations, we employed the Jordan-Wigner transformation~\cite{Jordan1928}. 
In addition, we opted for the restricted HF (RHF) formalism that further reduces the required computational resources.
For its implementation, we follow closely the steps and report the equations provided in the work of Elfving et al.~\cite{Elfving2020}. 
In essence, only half of the orbitals are required for the computation of the ground state of the systems considered in this paper, bringing the advantage of halving the total number of qubits. 
However, in this approach, the full Hilbert space is restricted only to the subspace of electronic states in which the orbitals are either empty or occupied by a pair of electrons (i.e. singlet states) limiting the applicability of this method.

The starting point is the RHF Hamiltonian that can be written as
\begin{equation}\label{eq:rhf_hamiltonian}
\hat{H}_{\text{RHF}}(\bm{R})=\sum_{rs} h_{rs}^{*}(\bm{R}) \hat{b}_{r}^{\dagger} \hat{b}_{s}+\sum_{r \neq s} g_{rs}^{*}(\bm{R}) \hat{b}_{r}^{\dagger} \hat{b}_{r} \hat{b}_{s}^{\dagger} \hat{b}_{s} + E_{NN}(\bm{R}),
\end{equation}
where, to the difference to its unrestricted version (Eq.~\eqref{eq:H}), the $\hat{b}_{r}^{\dagger}, \hat{b}_{r}$ are the creation and annihilation operators of a pair of electrons in the mode $r$.
More specifically, the indices $r$ and $s$ label the general (occupied or unoccupied) molecular orbitals.

This operator respects the following hard-core boson (anti-)commutation relations
\begin{equation}
\begin{aligned}
\left[\hat{b}_{r}, \hat{b}_{s}^{\dagger}\right]=\left[\hat{b}_{r}^{\dagger}, \hat{b}_{s}^{\dagger}\right]=\left[\hat{b}_{r}, \hat{b}_{s}\right] &=0 \quad(r \neq s), \\
\left\{\hat{b}_{r}^{\dagger}, \hat{b}_{r}^{\dagger}\right\}=&\left\{\hat{b}_{r}, \hat{b}_{r}\right\}=0, \\
&\left\{\hat{b}_{r}, \hat{b}_{r}^{\dagger}\right\}=1.
\end{aligned}
\end{equation}

The integrals $h_{rs}^{*}(\bm{R})$ and $g_{rs}^{*}(\bm{R})$ related to the original single-/two-electron integrals (see Eqs.~\eqref{eq:1eint} and~\eqref{eq:2eint}) as follows
\[
\begin{array}{l}
h_{rr}^{*}(\bm{R})=2 h_{rr}(\bm{R})+g_{rrrr}(\bm{R}), \\
h_{rs}^{*}(\bm{R})=g_{rrss}(\bm{R}) \quad(r \neq s), \\
g_{rs}^{*}(\bm{R})=2 g_{rssr}(\bm{R})-g_{rsrs}(\bm{R}) \quad(r \neq s).
\end{array}
\]

Using the fact that the operators $\hat{b}_{r}^{\dagger}$, $\hat{b}_{r}$
commute, they can be written using Pauli operators (see Eq.~\eqref{eq:pauli_spin_operators})
\begin{equation}
\hat{b}_{r}=\frac{1}{2}\left(\hat{\sigma}_{r}^{{x}}-i \hat{\sigma}_{r}^{{y}}\right),
\end{equation}
after the substitution into Eq.~\eqref{eq:rhf_hamiltonian}, the qubit Hamiltonian in Pauli representation reads as
\begin{equation}
\begin{aligned}
\hat{H}_{\text{RHF}}^{q} &= \sum_{r} \frac{h_{rr}^{*}(\bm{R})}{2}\left(\hat{I}_{r}-\hat{\sigma}_{r}^{z}\right) \\
&+\sum_{r \neq s} \frac{h_{rs}^{*}(\bm{R})}{4}\left(\hat{\sigma}_{r}^{{x}} \hat{\sigma}_{s}^{{x}}+\hat{\sigma}_{r}^{y} \hat{\sigma}_{s}^{{y}}\right) \\
&+\sum_{r \neq s} \frac{g_{rs}^{*}(\bm{R})}{4}\left(\hat{I}_{r}-\hat{\sigma}_{r}^{{z}}-\hat{\sigma}_{s}^{{z}}+\hat{\sigma}_{r}^{{z}} \hat{\sigma}_{s}^{{z}}\right) \\
&+ E_{NN}(\bm{R}).
\end{aligned}
\end{equation}

To summarize, for the simulations of the H$_2$ molecule we employ the parity transformation with two-qubit reduction, hence, the simulations require in total two qubits. 
Note that we do not consider for H$_2$ simulations to reduce to a single qubit problem as we want to evaluate the effect of noise that is predominantly due to two-qubit gates (CNOTs).
For H$_{3}^{+}$, we first employ the RHF formalism to have 3 qubit problem and then one qubit is `tapered off' reducing the problem also to two qubits. The use of the restricted formalism is justified by the fact that the dynamics only samples bonded molecular configurations.

\section{The RY Ansatz}\label{sec:appendixC_RYAnsatz}

For the implementation of the RY Ansatz, the gate $U3(\theta, \phi, \lambda)$ is defined as
\begin{align}\begin{aligned}\begin{split}U3(\theta, \phi, \lambda) =
    \begin{pmatrix}
        \cos(\frac{\theta}{2})          & -e^{i\lambda}\sin(\frac{\theta}{2}) \\
        e^{i\phi}\sin(\frac{\theta}{2}) & e^{i(\phi+\lambda)}\cos(\frac{\theta}{2})
    \end{pmatrix}\end{split},\end{aligned}\end{align}
and the gate $U2(\phi, \lambda)$ is written as
\begin{align}\begin{aligned}\begin{split}U2(\phi, \lambda) = \frac{1}{\sqrt{2}}
    \begin{pmatrix}
        1          & -e^{i\lambda} \\
        e^{i\phi} & e^{i(\phi+\lambda)}
    \end{pmatrix}\end{split}\end{aligned}\end{align}
where $\theta, \phi, \lambda$ are the Euler angles.

\section{Extended dynamics of H$_2$}\label{sec:appendixE_longverlet}

\begin{figure}[ht]
\centering
\includegraphics[width=1\linewidth]{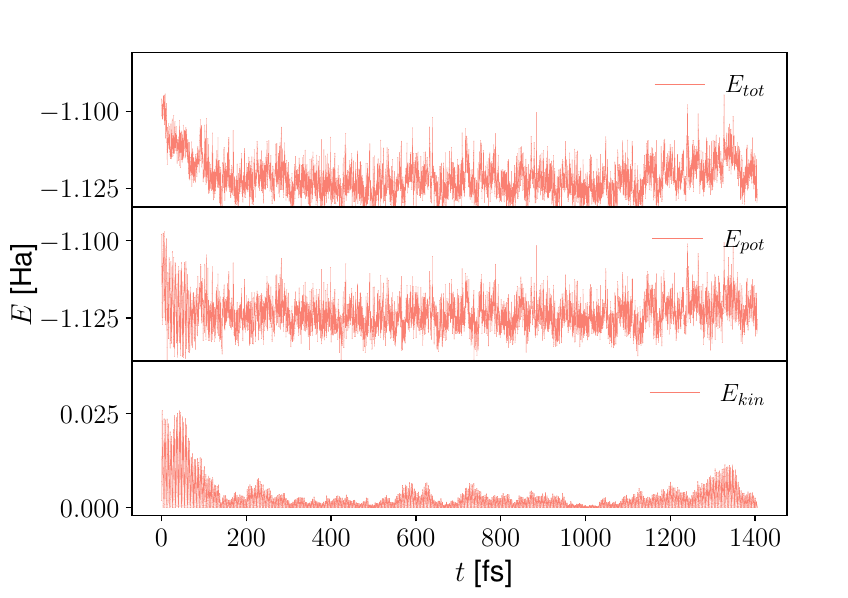}
\caption{{
Time series of the total, potential and kinetic energies for the dynamics of the H$_2$ molecule ($t = 1400$ fs, $dt= 0.2$ fs) using the VQE algorithm with the realistic noise corresponding to the \textit{ibmq\_athens} device and 8'192 measurements for the evaluation of energies and forces.
}}

\label{fig:h2_verlet_1200fs}
\end{figure}

To demonstrate the behaviour of microcanonical dynamics beyond 100 fs in a noisy setting, we extend the duration to 1400 fs and show the results in Figure~\ref{fig:h2_verlet_1200fs}. 
We use the standard parameters (i.e. same settings as for the Fig.~\ref{fig:h2_all_in_one}(a))  defined in Section~\ref{sec:results}.

\bibliographystyle{elsarticle-num}
%\bibliography{bibliography}

\begin{thebibliography}{10}
\bibitem{notemass}
The notation ${\bm X } / {\bm m}$, where $\bm X$ is a $3N$ dimensional vector
  like the velocity $\bm v$ or the force $\bm F$ vector, implies that the
  component ${X}_{I \alpha}$ of the vector${\bm X}$ corresponding to the $I$-th
  ion and $\alpha$-th direction, is divided by the mass $m_I$ of the $I$-th
  ion.
\end{thebibliography}

\end{document}